\documentclass[11pt, a4paper]{article} 
\pdfoutput=1
\usepackage[utf8x]{inputenc}     % if unicode errors are coming
\usepackage{jheppub}
\usepackage{enumerate}
\usepackage{epsfig} 
\usepackage{float} 
\usepackage[caption = false]{subfig}
\usepackage{wasysym} 
\usepackage{mathrsfs} 
\usepackage{amsfonts} 
\usepackage{amsbsy} 
\usepackage{amscd} 
\usepackage{cancel}
\usepackage{pstricks} 
\usepackage{multirow} 
\usepackage{tikz}
\usepackage{color}
\usepackage{xcolor}
\usepackage{slashed}
\usepackage{array}
\usepackage{tablefootnote}
\usepackage[normalem]{ulem}

\usetikzlibrary{arrows,positioning,shapes.geometric} 
\usepackage[font=small,labelfont=bf]{caption}
\usepackage{diagbox}
\usepackage{multirow}
\usepackage{tabularx}
\usepackage{soul}  % Strikethrough text with \st{Hellow world}
\usepackage{listings} 
\usepackage{cleveref}
\usepackage{color}

\definecolor{somethingC}{rgb}{0.07, 0.04, 0.56} 
\usepackage{amsthm}

\newcommand{\stkout}[1]{\ifmmode\text{\sout{\ensuremath{#1}}}\else\sout{#1}\fi}
\title{Unveiling desert region in inert doublet model assisted by Peccei-Quinn symmetry}
\author[]{Anupam Ghosh}
\author[]{and Partha Konar}

\affiliation[]{Theoretical Physics Division, Physical Research Laboratory, Shree Pannalal Patel Marg, Ahmedabad, 380009, Gujarat, India}

\emailAdd{anupam@prl.res.in}
\emailAdd{konar@prl.res.in}

\abstract{
The Inert Higgs Doublet model (IDM), assisted by Peccei-Quinn (PQ) symmetry, offers a simple but natural framework of a dark sector that accommodates Weakly Interacting Massive Particle (WIMP) and axion as dark matter components. Spontaneous breaking of $U(1)_{PQ}$ symmetry, which was originally proposed as an elegant solution to the strong charge-parity (CP) problem, also ensures the stability of WIMP through a residual $\mathbb{Z}_2$ symmetry. 
Interestingly, additional fields necessitated by PQ symmetry further enrich the dark sector. These include a scalar field proprietor for axion DM and a vector-like quark (VLQ) that acts as a portal for the dark sector through Yukawa interactions.
Moreover, this combination of the axion and WIMP components satisfies the observed DM relic density and reopens the phenomenologically exciting region of the IDM parameter space where the WIMP mass falls between 100 - 550 GeV.
We investigate the model-independent pair production of VLQs exploring this region at the Large Hadron Collider (LHC), incorporating the effects of next-to-leading order (NLO) QCD corrections. After production, each VLQ decays into a top or bottom quark accompanied by an inert scalar, a consequence of the residual $\mathbb{Z}_2$ symmetry.
Utilising relevant observables with a leptonic search channel and employing multivariate analysis, we demonstrate the ability of this analysis to exclude a significant portion of the parameter space with an integrated luminosity of 300 $\text{fb}^{-1}$. 
}
\preprint{\today}

\begin{document}
\maketitle
%\flushbottom

\newpage

%======================================================================================
\section{Introduction}
\label{intro}
%======================================================================================
%%%%%%%%%%%%%%%%%%%%%%%%%

The Standard Model (SM) of particle physics has established itself as a cornerstone of modern physics, remarkably successful in describing a vast array of experimental data during the last several decades. However, its explanatory reach remains limited in the face of several fundamental enigmas, such as the strong charge-parity (CP) problem \cite{Peccei:1977hh, Peccei:1977ur}, the existence of a substantial component of non-baryonic dark matter (DM) in the Universe \cite{Sofue:2000jx, Clowe:2006eq}, the tiny yet non-zero masses of neutrinos \cite{Super-Kamiokande:1998kpq, SNO:2002tuh, K2K:2002icj}, and the observed matter-antimatter asymmetry \cite{Riotto:1999yt, Dine:2003ax}, among others.
These persistent shortcomings motivated the theoretical and experimental communities to explore new physics theories beyond the standard model (BSM), extending the SM's explanatory power. This work delves into a simple yet elegant extension of the SM that tackles two of these outstanding challenges: the Strong CP problem and the existence of dark matter. We propose a detailed search strategy at the Large Hadron Collider (LHC) to probe the associated dark sector particles.

Inert Higgs doublet model (IDM) \cite{Barbieri:2006dq, Cirelli:2005uq} provides one of the simplest BSM frameworks with a viable dark matter candidate. The standard model is extended by a single $SU(2)_L$ scalar doublet carrying an odd charge under $\mathbb{Z}_2$ symmetry, resulting in three new states: charged scalar ($H^\pm$), a neutral scalar ($H^0$), and a neutral pseudoscalar ($A^0$). Notably, the lightest of these neutral components become stable and a dark matter candidate since an ad hoc $\mathbb{Z}_2$ symmetry is implemented to prevent the state from decaying.
Various experimental searches for dark matter and theoretical considerations strongly constrain this extended DM model. 
Direct detection experiments,  LHC searches, and other astronomical and cosmological observations constrain the model's parameter space \cite{Ilnicka:2015jba, Belyaev:2016lok, Arhrib:2013ela, Dercks:2018wch}. Theoretical constraints also play a crucial role. Oblique electroweak parameters ($S, T, U$) \cite{Peskin:1991sw} quantify contributions from BSM physics to electroweak radiative corrections. Notably, to satisfy the $T$-parameter constraint, the masses of two out of the three new IDM particles often need to be nearly degenerate \cite{Belyaev:2016lok}.
The IDM can explain the measured dark matter relic density within two characteristically distinct parameter regions, such as the hierarchical mass region and the degenerate region. The hierarchical mass scenario achieves this through resonant Higgs portal annihilation with the dark matter mass roughly half that of the SM Higgs boson, while the other two BSM scalars ($H^\pm, A^0$) are heavier and almost degenerate. 
In the degenerate region, where all three new particles have nearly equal masses, the dark matter typically has a mass exceeding 550 GeV. 
Consequently, the relic density remains underabundant due to strong gauge interactions for a significant parameter space where dark matter mass lies between 100-550 GeV.
The prospect of searching for IDM particles at the LHC is particularly interesting. For example, hierarchical regions can be probed through the signature of jets or boosted fatjets plus missing transverse momentum (MET) \citep{Ghosh:2021noq,Belyaev:2018ext,Poulose:2016lvz}. Displaces vertex search \cite{CMS:2014gxa} can offer a potential avenue for exploring the degenerate mass region. However, one expects a more challenging pursuit due to the heavy final state particles and their ultra-soft decay products.

Non-barionic cold dark matter, constituting roughly $27\%$ of the universe's energy budget, remains a mystery. While a simple, single-candidate model holds appeal, no such supportive evidence exists so far. On the other hand, the universe's visible matter and radiation components, comprising only a tiny fraction ($5\%$) of energy budget, are composed of a spectrum of particles from fermionic families, gauge bosons and Higgs. Such design insinuates the possibility of a much richer structure within the dark sector, including a multicomponent dark matter scenario.
The WIMP-axion scenario presents an exciting potential, not only because the combined contributions with axion can yield the observed dark matter relic density. Additionally, spontaneous breaking of Peccei-Quinn $U(1)_{PQ}$ symmetry provides an elegant solution to the strong CP problem and also ensures the stability of WIMP through a residual $\mathbb{Z}_2$ symmetry for a natural realisation of this dark matter picture. 

Spontaneous PQ symmetry breaking necessitates a complex scalar field, with its phase component being the axion. This $U(1)_{PQ}$ symmetry is expected to break at an energy scale considerably higher than the electroweak scale. Spontaneous breaking of this symmetry results in the emergence of a pseudo-Goldstone boson, commonly known as the QCD axion. This axion possesses faint interactions with all other particles \cite{Kim:1979if, Shifman:1979if, Dine:1981rt, Zhitnitsky:1980tq} and can have a lifetime far exceeding the universe's age, making it a suitable dark matter candidate. The PQ symmetry breaking scale and the misalignment angle determine the axion relic density.
To implement the Peccei-Quinn symmetry, a vector-like quark (VLQ) is necessary \cite{Peccei:1977hh, Peccei:1977ur}. 
The vector-like quark facilitates a straightforward realisation of the $U(1)_{PQ}$ symmetry, as seen in the KSVZ axion model \cite{Kim:1979if, Shifman:1979if} and serves as a portal between the dark sector and the SM fields. 
This connection is evident by the spontaneous breaking of the $U(1)_{PQ}$ symmetry that generates a residual $\mathbb{Z}_2$ symmetry inherently stabilising the WIMP dark matter candidate~\footnote{Without any loss of generality, we can further continue our analysis considering $H^0$ as the lightest mode and the WIMP DM candidate.}, thus eliminating the need for the introduction of an additional ad-hoc discrete symmetry in the model.
Furthermore, the two-component IDM framework reopens the intriguing 100-550 GeV mass window for WIMPs and presents a compelling avenue for further exploration of dark matter.

Building upon the WIMP-axion scenario, the residual $\mathbb{Z}_2$ symmetry in the inert Higgs doublet model dictates that the vector-like quark ($\Psi$) decays exclusively into BSM scalars ($H^\pm, A^0, H^0$) with SM quark. Recent study \cite{Alves:2016bib} has explored the IDM with PQ symmetry in detail, focusing on dark matter phenomenology. Their analysis considered a scenario where the VLQ interacts only with the first-generation quark family of the SM. Additionally, they reinterpreted LHC mono-jet and multi-jet and missing-energy searches, demonstrating the feasibility of a VLQ-SM quark coupling strength ranging from 0 to 1 across various parameter spaces.
In contrast, the present work focuses on VLQs primarily interacting with the third-generation quark family (top and bottom quarks) of the SM, as the Yukawa interactions in the visible sector. Interactions with the first and second generations are considered negligible. This approach leads to a significant advantage: the VLQ pair production cross-section becomes independent of any BSM couplings, relying solely on the strong coupling constant. This independence renders our searches model-independent.
Furthermore, this study identifies an exciting signal topology at the LHC for exploring the degenerate region of the IDM parameter space. This signal consists of a single, isolated electron or muon accompanied by two jets (one of which is identified as a b-tagged jet) with significant missing transverse momentum (MET).

Hence, the salient features of this work are as follows:
(i) The present work explores the capabilities of the LHC to probe the WIMP-axion multicomponent dark matter scenario within the Inert Higgs doublet framework, where the residual discrete symmetry inherent to the model naturally stabilises the WIMP dark matter candidate.
(ii) Our study focuses on the phenomenologically interesting degenerate region ($100~\text{GeV} \leq M_{H^0} \leq 700~\text{GeV}$) of the IDM parameter space, considered as a desert region in a pure IDM scenario. 
(iii) Traditionally, this heavier DM region in IDM (degenerate region) is challenging to probe at the LHC because the heavy particles ($H^\pm, A^0$) decay into the dark matter particle ($H^0$) along with low-momentum Standard Model particles. However, the present scenario opened up a novel search strategy that leverages the presence of vector-like quarks within the model. This strategy targets an alternative signal topology consisting of an isolated, energetic lepton (electron or muon) accompanied by two jets (one identified as a b-jet) and significant missing transverse momentum. 
(iv) We demonstrate that a sizeable portion of the degenerate parameter space can be explored using this approach at the 14 TeV LHC already with 300 fb$^{-1}$ of luminosity.
(v) To enhance the accuracy of our predictions, we incorporate Next-to-Leading-Order (NLO) QCD corrections for the VLQ pair production process. 
(vi) Additionally, we utilise a sophisticated multivariate analysis technique employing a Boosted Decision Tree (BDT) algorithm to refine the collider analysis further.

The rest of the manuscript is organised as follows: Section \ref{theory} discusses the underlying theoretical framework of the model. We provide a concise overview of various constraints arising from theoretical considerations, astrophysical observations, and existing collider data. This section also explores the dark matter phenomenology associated with the two-component dark matter scenario. In the next Section \ref{LHC-pheno}, we delve into the collider phenomenology, describing the targeted signal topologies and establishing suitable benchmark points for our analysis. We detail the primary event selection process based on simulated signal and background events, followed by constructing relevant high-level variables. Subsequently, we employ multivariate analysis techniques on these input variables to calculate the corresponding signal efficiency. Finally, this section presents the final statistical significance for different benchmark points and establishes exclusion regions within the parameter space. Finally, the concluding Section \ref{conc} summarises our key findings and reiterates the potential of our proposed search strategy for exploring the degenerate region of the inert doublet model assisted by Peccei-Quinn symmetry at the LHC.

%===========================================================================================================================
\section{Theoretical Framework}
\label{theory}
%======================================================================================
\subsection{The model}
\label{model}
%======================================================================================
The model comprises an inert Higgs doublet model augmented by a KSVZ-type axion model \cite{Kim:1979if, Shifman:1979if}. This axion model includes a singlet scalar field, with its phase part being the axion field. The vacuum-expectation value (vev) of this field breaks the $PQ$ symmetry at a scale much larger than the electroweak (EW) scale. A residual $\mathbb{Z}_2$ symmetry will remain intact after $PQ$ symmetry breaking, stabilising the WIMP dark matter, the lightest neutral component of the IDM. The scalar field is as described below.
\begin{equation}
\eta = \frac{1}{\sqrt{2}} \Bigl(f_a+\sigma(x)\Bigr)~e^{\frac{ia(x)}{f_a}}.
\label{Eq.eta}
\end{equation}
The $PQ$ breaking scale is denoted as $f_a$ and corresponds to vev of $\eta$, while $\sigma(x)$ and $a(x)$ represent the radial mode and axion field. The axion is invariably associated with $\frac{a(x)}{f_a}$; its interactions with matter and gauge bosons are diminished by $f_a$. Since $f_a$ is much larger than the EW scale, axions have faint interactions with other particles. The effects of nonperturbative QCD result in a potential that gives a mass to the axion \cite{GrillidiCortona:2015jxo}. 
\begin{equation}
m_a = 5.70 ~\Bigl( \frac{10^{12}~\text{GeV}}{f_a} \Bigr)~\mu eV~.
\end{equation}

In addition to the Standard Model Higgs doublet, $\Phi_1 \sim (1,2,1/2)$, IDM contains a new $SU(2)_L$ doublet, $\Phi_2 \sim (1,2,1/2)$. The numbers inside the parenthesis represent the charges under the SM gauge groups $SU(3)_c \otimes SU(2)_L \otimes U(1)_Y$. Peccei-Quinn charge of $\Phi_2$ is $PQ(\Phi_2)=-1$, while all the SM fields have zero $PQ$ charges. Fields with even or zero $PQ$ charge remain even under residual $\mathbb{Z}_2$ symmetry after $PQ$ symmetry breaking, while fields with odd $PQ$ charge become odd under $\mathbb{Z}_2$. The two doublets can be written as
\begin{equation}
\Phi_1= \begin{pmatrix}
G^+\\
\dfrac{1}{\sqrt{2}}(v+h+iG^0)
\end{pmatrix}
, 
\quad\Phi_2= \begin{pmatrix}
H^+\\
\dfrac{H^0 + i \hspace{1mm} A^0}{\sqrt{2}}
\end{pmatrix},
\end{equation}
where $G^0$ and $G^+$ are the Goldstone bosons, $h$ is the SM Higgs boson and $v=246 \hspace{1mm} \text{GeV}$ is the EW vev. $H^\pm$, $H^0$, and $A^0$ are the inert scalars, with $H^0$ and $A^0$ being CP even and CP odd neutral scalars, respectively. The Higgs potential is as follows,
\begin{equation} 
\begin{split}
V  = & \, \mu_1^2\Phi_1^\dagger\Phi_1+ \mu_2^2\Phi_2^\dagger\Phi_2+ \dfrac{\lambda_1}{2} (\Phi_1^\dagger\Phi_1)^2+ \dfrac{\lambda_2}{2} (\Phi_2^\dagger\Phi_2)^2   
+ \lambda_3 (\Phi_1^\dagger\Phi_1)(\Phi_2^\dagger\Phi_2)  + \lambda_4 (\Phi_2^\dagger\Phi_1)(\Phi_1^\dagger\Phi_2) \\ &
+ \dfrac{\lambda_5}{2}\hspace{1mm} [(\Phi_1^\dagger\Phi_2)^2+(\Phi_2^\dagger\Phi_1)^2] \, . 
\end{split}
\label{Eq:IDM_pot}
\end{equation}
After EW symmetry breaking, the masses for inert scalars are as follows.
\begin{equation}
\begin{split}
&  M^2_{H^\pm}=\mu_2^2+\dfrac{1}{2}\lambda_3 v^2,   \\ &
M^2_{A^0} =\mu_2^2+\dfrac{1}{2}\lambda_c v^2,  \\ &
 M^2_{H^0} =\mu_2^2+\dfrac{1}{2}\lambda_L v^2 .
\end{split}
\label{Eq:mass_inert}
\end{equation}
where $\lambda_{L/c}=(\lambda_3+\lambda_4\pm \lambda_5)$, $\lambda_{L}$ is also known as a Higgs portal coupling and can be either positive or negative. The parameters $\lambda_1$ and $\mu_1$ can be expressed in terms of the mass of the SM Higgs boson ($M_h=125$ GeV) and the EW vev ($v$). $\lambda_2$ represents the self-coupling strength between inert scalars, is a free parameter, and has no effect on scalar masses and their phenomenology. In our study, we consider $H^0$ to be the dark matter candidate. Alternatively, one can choose $A^0$ as the dark matter candidate without altering any phenomenology simply by reversing the sign of $\lambda_5$. 

Addressing the strong CP problem, the KSVZ axion model requires the inclusion of at least one heavy vector-like quark ($\Psi$) in addition to the SM fermions. The VLQ has the following interaction with the PQ breaking scalar ($\eta$)
\begin{equation}
f_\Psi ~ \eta^{*}~ \overline{\Psi}_L \Psi_R + h.c.~,
\end{equation}
where $PQ(\Psi_L)=-1$ and $PQ(\Psi_R)=+1$ and $f_\Psi$ is the coupling strength. The above equation facilitates the interaction of axions with a pair of color triplet VLQs, enabling the axions to interact indirectly with gluons via a triangle-loop diagram with VLQ in the loop. This axion-gluon field strength interaction is necessary to solve the strong CP problem through the $PQ$ mechanism. The charge assignments of all BSM fields are provided in table \ref{tab:charge}.
%==================
\begin{table}[tb!]
\begin{center}
%\scriptsize
 \begin{tabular}{|c|c|c|c|c|}
\hline
 & $\eta$ & $\Psi_L$ & $\Psi_R$ & $\Phi_2$   \\
\hline\hline
 $SU(3)_C$ & 1 & 3 & 3 & 1 \\  
\hline
 $SU(2)_L$ & 1 & 1 & 1 & 2\\  
\hline
 $U(1)_{PQ}$ & 2 & -1 & 1 & -1 \\  
\hline
 $\mathbb{Z}_2$ & + & - & - & -  \\  
\hline
 \end{tabular} 
\caption{Particle contents beyond the SM and their quantum charges.}
\label{tab:charge}
\end{center}
\end{table}
%==================
VLQ bridges dark sector particles and SM fields, featuring the following Yukawa interaction with the SM quark doublet $q_L \sim (3,2,1/6)$ and the IDM doublet $\Phi_2$.
\begin{equation}
\mathcal{L} \supset f ~\bar{q}_L \Phi_2 \Psi_R + h.c.
\label{Eq:yukawa}
\end{equation}
The $U(1)_Y$ hypercharge of VLQ is $-1/3$ and remains odd under residual $\mathbb{Z}_2$ symmetry after the breaking of $PQ$ symmetry. As a result, VLQ does not mix with the SM quarks. VLQ can interact with all three SM quark generations. However, interactions primarily with the first two generations are heavily constrained by flavor-violating effects, such as $D^0-\bar{D}^0$ oscillation \cite{Garny:2014waa, Gedalia:2009kh}. To be consistent with flavour oscillation data, interactions of the VLQ must predominantly involve a single quark family. Another exciting possibility is that the TeV scale VLQ predominantly interacts with the third generation of SM quarks and minimally (or not at all) with the first two generations, similar to the SM Higgs boson coupling with the top quark being much larger compared to the up and charm quark couplings. In this case, the interaction vertices from equation \ref{Eq:yukawa} are $\Psi \bar{t} H^{+}$, $\Psi \bar{b} H^{0}$, and $\Psi \bar{b} A^{0}$. The free parameters of the model are the following.
\begin{equation}
\bigl\{M_{H^0}, M_{A^0}, M_{H^\pm}, M_{\Psi}, \lambda_L, f, f_a \bigr\}
\label{Eq:free}
\end{equation}

The mass of VLQ is $M_{\Psi}=f_\Psi f_a/\sqrt{2}$. The coupling $f_\Psi$ becomes the dependent parameter since we consider $M_{\Psi}$ and the axion decay constant ($f_a$) independent parameters. All the inert scalars have nearly degenerate mass in the degenerate region, $\Delta M=M_{H^\pm}-M_{H^0}=M_{A^0}-M_{H^0}=\mathcal{O}(1)$ GeV.  Pair production of vector-like quarks in our analysis does not depend on the $\lambda_L$ coupling. However, the same does play a role in the IDM dark matter phenomenology of $H^0$. Various theoretical, astrophysical, and collider data can constrain this model, and we will discuss them briefly in the next.\\

In article \cite{Chatterjee:2018mac}, the authors studied the WIMP-axion multi-component dark matter framework. They considered a KSVZ model augmented with a complex scalar, which is singlet under SM gauge groups but has a non-zero charge under $U(1)_{PQ}$. The model was examined in detail, considering constraints from various experiments such as dark matter direct detection, indirect detection, and LHC data. In the model, vector-like quark (VLQ) has Yukawa interactions with the down-type SM quarks and WIMP dark matter. The authors also did a signal background analysis and constrained the model in the coupling versus DM mass (WIMP) plane in the $40~\text{GeV} \leq M_{\text{DM}} \leq 100~\text{GeV}$ range from the dijet+MET and monojet+MET searches.
 In contrast, we have an inert doublet that is an $SU(2)_L$ doublet and has a non-zero $U(1)_{PQ}$ charge. Additionally, the VLQ in our model has Yukawa interaction with the SM quark doublet and the IDM doublet, resulting in VLQ decay into both up-type and down-type SM quarks along with the inert scalars. Due to the interaction with up-type SM quark ($\Psi \bar{t} H^{+}$), an exciting signal topology is observed to search our multi-component dark matter model at the LHC. Our collider analysis mainly focuses on the region $M_{\text{DM}} \geq 100~\text{GeV}$.

%======================================================================================
\subsection{Constraints}
\label{Constraints}
%======================================================================================
For a realistic model to be considered valid, the potential must be bounded from below, and the vacuum must be charge-neutral, which leads to the following constraints \cite{Deshpande:1977rw, Ivanov:2006yq}: 
\begin{equation}
 \lambda_1 > 0, \hspace{1mm} \lambda_2 > 0, \hspace{1mm} \lambda_3 + 2\sqrt{\lambda_1 \lambda_2} > 0, 
 \hspace{1mm} \lambda_3 + \lambda_4 + \lambda_5 +  2\sqrt{\lambda_1 \lambda_2} > 0.
\end{equation}
 The condition $\lambda_4 + \lambda_5 < 0$ implies that the inert vacuum is charge neutral. The requirement of the inert vacuum to be the global minima leads to \cite{Ginzburg:2010wa, Swiezewska:2012ej},
\begin{equation}
\dfrac{\mu_1^2}{\sqrt{\lambda_1}} - \dfrac{\mu_2^2}{\sqrt{\lambda_2}} > 0 ~.
\end{equation}
The unitarity in $2 \rightarrow 2$ scalar-scalar scattering processes demands that all $|\Lambda_i| \leq 8\pi$, where $|\Lambda_i|$ are given below \cite{Arhrib:2012ia}.
\begin{equation}
\begin{split}
& \Lambda_{1,2} = \lambda_3 \pm \lambda_4, \hspace{1mm} \Lambda_{3,4} = \lambda_3 \pm \lambda_5, \hspace{1mm} \Lambda_{5,6} = \lambda_3 + 2 \lambda_4 \pm 3\lambda_5, \\&
\Lambda_{7,8} = -\lambda_1 - \lambda_2 \pm \sqrt{(\lambda_1- \lambda_2)^2 + \lambda_4^2}, \hspace{1mm}\\& 
\Lambda_{9,10} = -3 \lambda_1 - 3 \lambda_2 \pm \sqrt{9(\lambda_1- \lambda_2)^2 + (2\lambda_3 + \lambda_4)^2}, \\&
\Lambda_{11,12} = -\lambda_1 - \lambda_2 \pm \sqrt{(\lambda_1- \lambda_2)^2 + \lambda_5^2} .
\end{split}
\end{equation}
The SM weak gauge bosons can not decay into dark scalars, constraining the inert scalars' masses.
\begin{equation}
M_Z < M_{H^0}+ M_{A^0},~ M_{W^\pm} < M_{H^0}+ M_{H^\pm},~ M_{W^\pm} < M_{A^0}+ M_{H^\pm}~.
\label{Eq:Decay_const}
\end{equation}
In the region where $M_{H^0} < \dfrac{M_h}{2}$, the SM Higgs boson can decay into a pair of WIMPs ($H^0$), contributing to the Higgs invisible decay branching ratio and imposing stringent limits on the Higgs portal coupling. For instance, if $M_{H^0}= 60~ \text{GeV} ~(10~\text{GeV})$, then $|\lambda_L|\lesssim 0.012 ~(0.007)$ \cite{Belanger:2013xza}.
\paragraph{LEP bound:} Supersymmetric neutralino search at LEP \cite{Lundstrom:2008ai, Belanger:2015kga} excludes the following region of the parameter spaces $M_{H^0}<80~\text{GeV},~ M_{A^0} <100~\text{GeV}$ for $M_{A^0}-M_{H^0} >8~\text{GeV}$. The supersymmetric chargino search \cite{Pierce:2007ut} at LEP further excludes $M_{H^\pm}<70~\text{GeV}$. Thus, we have chosen, 
\begin{equation}
M_{A^0},M_{H^\pm} > 100~\text{GeV},~\text{and}~M_{H^0} > 90~\text{GeV}~.
\label{Eq:search}
\end{equation}
\paragraph{LHC bound:} The authors \cite{Belanger:2015kga} analyzed the dilepton plus missing energy signature of LHC Run I data and excluded $H^0$ masses below 35 GeV at a $95\%$ confidence level (CL). This constraint was based on the production channel $q \bar{q}\rightarrow Z \rightarrow A^0 H^0\rightarrow (Z^{*}H^0) H^0\rightarrow l^+ l^- H^0 H^0 $. Our region of interest, equation \ref{Eq:search}, remains unaffected by this search. 

In a recent study conducted by the ATLAS collaboration \cite{Marjanovic:2014eca}, researchers searched for supersymmetric particles with a final state of multiple jets plus missing energy. Since vector-like quarks can interact with the first two generations of SM quarks, they can produce the same final state. The green shaded area in figure \ref{fig:result} represents the excluded parameter space at a $95\%$ CL based on the reinterpretation \cite{Giacchino:2015hvk} of the ATLAS results \cite{Marjanovic:2014eca} at the 8 TeV LHC, with an integrated luminosity of 20.3 $\text{fb}^{-1}$.

%======================================================================================
\subsection{Direct detection} 
\label{DD}
%======================================================================================
WIMPs may interact with nuclei, causing detectable energy deposits in underground instruments such as LUX \cite{LUX:2015abn}, XENON1T \cite{XENON100:2012itz}, and CDMS \cite{SuperCDMS:2014cds}. Each instrument uses different target nuclei, such as xenon, argon, and others, and unique detection methods. These underground detectors in direct detection experiments enforce strict constraints on the WIMP-nucleon scattering cross-section versus the WIMP mass.

The neutral inert scalars ($H^0, A^0$) interact with the $Z$ boson (interaction vertex is $H^0 A^0 Z$). In the case of degenerate $H^0$ and $A^0$, the spin-independent cross-section between the DM candidate ($H^0$) and the nucleon exceeds the current limits \cite{CDMS:2008uih,Alner:2007ja}. A mass difference order of 100 keV between the CP even ($H^0$) and CP odd ($A^0$) scalars is needed to circumvent this constraint \cite{Barbieri:2006dq, T.Hambye_2009}. This mass splitting kinematically suppresses the DM-nucleon interaction through a tree-level $Z$-boson exchange. For pure IDM, the constraints posed by the LUX \cite{Blinov:2015qva} and the expected sensitivity of the XENON1T \cite{Aprile:2012zx, XENON:2015gkh} can be circumvented by ensuring that the mass difference between $H^0$ and $A^0$ is greater than 100 keV and $|\lambda_L|\lesssim 0.01$. However, larger values of $\lambda_L$ are permissible in our multi-component dark matter framework as the WIMP-nucleon scattering cross-section is rescaled by a factor of $\frac{\Omega_{H^0}h^2}{\Omega_{\text{total}}h^2}$ (specified in equation \ref{Eq:tot_relic}). For instance, in figure \ref{Fig:relic}, a higher value of $\lambda_L$ leads to a smaller contribution from the WIMP in the relic density, making it consistent with the constraints from direct detection.

 The presence of an exotic vector-like quark ($\Psi$) can mediate the interaction of WIMP with nuclei through $\Psi$-mediated scattering $q H^0 \rightarrow q H^0$. The WIMP-nucleon scattering cross-section will be negligibly small if the $\Psi$ predominantly couples with the third-generation quark family. However, if the $\Psi$ couples with the first two generations of SM quarks, the cross-section can be significant and depends on the Yukawa coupling ($f$) and VLQ mass ($M_\Psi$). The most stringent constraint will occur when $M_\Psi \sim M_{H^0}$. If $f \lesssim 0.5$, the model will be consistent with the LUX constraint on the spin-independent scattering cross-section. If the mass of $\Psi$ is not degenerate to the WIMP ($H^0$), the bound will be much more relaxed. This holds true if $H^0$ contributes to all of the dark matter relic density. However, this is not the case in our two-component dark matter scenario. The WIMP-nucleon cross-section needs to be rescaled by the ratio of WIMP local density to the total relic density. As a result, the constraints will be further relaxed in cases where the axion contribution is significant.

%======================================================================================
\subsection{Indirect detection}
\label{ID}
%======================================================================================
The indirect detection concept involves observing particles produced when dark matter annihilates or co-annihilates, generating final state particles like electrons, positrons, photons, and neutrinos. In many BSM scenarios, dark matter particles annihilate to produce SM particles. Photons and neutrinos, being electromagnetically neutral, are likely to reach detectors without much deviation. Experiments such as PAMELA \cite{Kohri:2009yn}, MAGIC \cite{MAGIC:2016xys} and Fermi-LAT \cite{Eiteneuer:2017hoh}, among others, are searching for these indirect detection signals. By measuring the flux of these stable particles, constraints can be placed on the parameters of various dark matter models. For a two-component dark matter scenario, the indirect detection cross-section needs to be rescaled by the ratio of the local density of $H^0$ to the total dark matter relic density squared. As a result, the mass of interest $100~\text{GeV} < M_{H^0} < 550~\text{GeV}$ is consistent with the searches by Fermi-LAT \cite{Eiteneuer:2017hoh} and H.E.S.S. \cite{HESS:2011zpk} limits.

%======================================================================================
\subsection{Dark matter relic density}
\label{relic}
%======================================================================================
%==================
\begin{figure}[tbh!]
\centering
\includegraphics[scale=0.6]{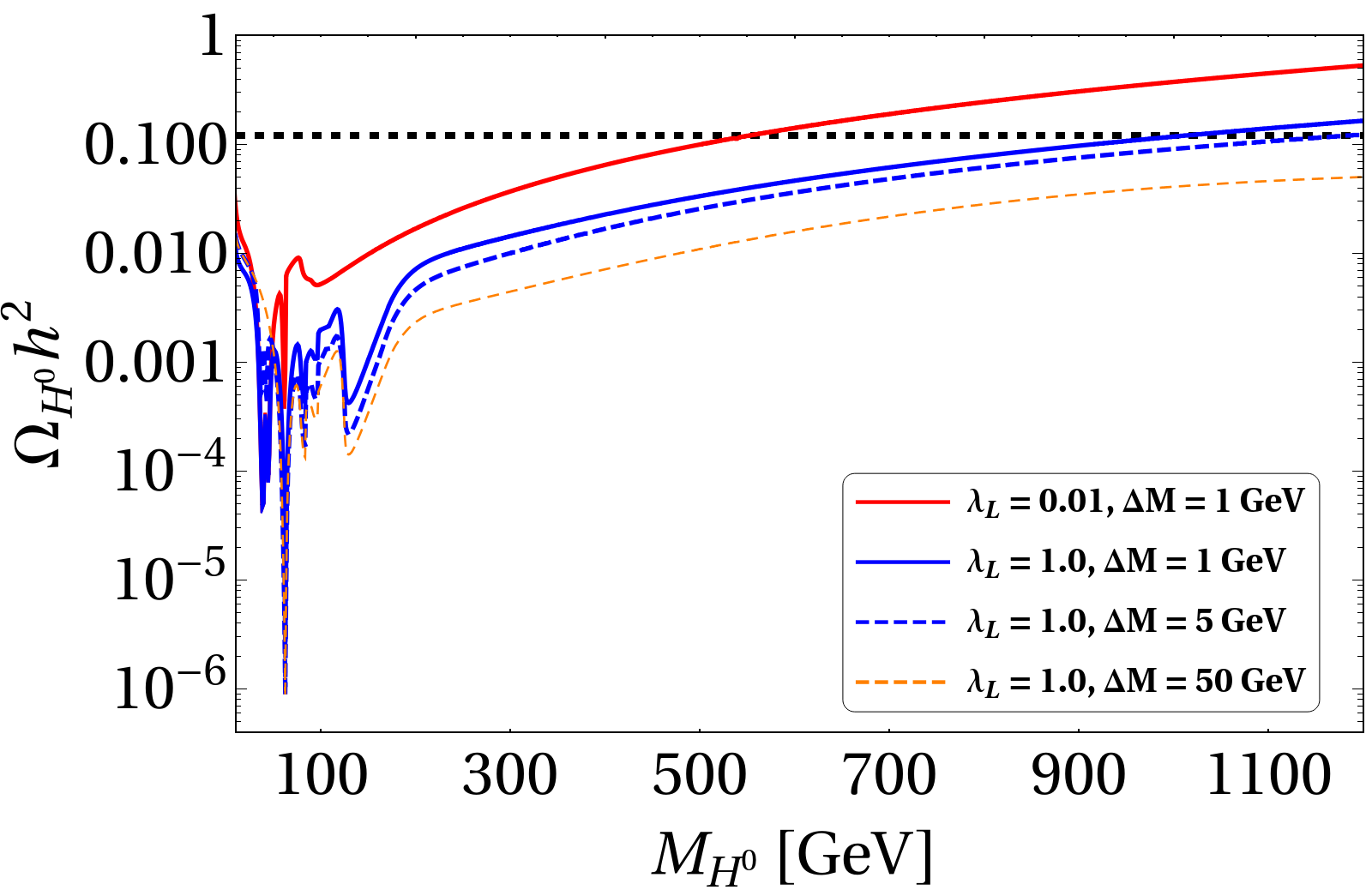}
\caption{The plot shows the variation of WIMP relic density for different DM masses when inert scalars are near mass degenerate, but VLQ is not. The red line represents the Higgs portal coupling with $\lambda_L=0.01$ and $\Delta M=1$ GeV. The solid (dashed) blue line is for $\lambda_L=1.0$ and $\Delta M=1$ GeV (5 GeV). Here, $\Delta M=M_{H^\pm}-M_{H^0}=M_{A^0}-M_{H^0}$. The dashed black line is the observed DM relic density by the Planck collaboration \cite{Aghanim:2018eyx}.}
\label{Fig:relic}
\end{figure}
%==================

In figure \ref{Fig:relic}, we display the WIMP ($H^0$) abundance, considering VLQ is not degenerate with the inert scalars. The relic abundance is obtained by solving the Boltzmann equation using \texttt{micrOMEGAs} -v5 \cite{Belanger:2018ccd}. As mentioned earlier, in IDM, there are two possible regions: (1) Hierarchical, where $A^0$ and $H^\pm$ have nearly identical masses significantly higher than $H^0$ mass, and (2) Degenerate, where all inert scalar masses are nearly identical, allowing for coannihilation. In the former case, no coannihilation is present because of the larger mass gap between heavy inert scalars and $H^0$. In that case, the correct relic is obtained through annihilation of $H^0$ into the SM Higgs boson in the region when the mass of $H^0$ is nearly half of the SM Higgs boson mass. The exploration of the hierarchical region at the LHC has been extensively documented in the literature \cite{Ghosh:2021noq, Poulose:2016lvz, Datta:2016nfz}. The variation of the relic density of $H^0$  in figure \ref{Fig:relic} is shown for different values of the Higgs portal coupling $\lambda_L=0.01~\text{and}~1.0$, keeping $\Delta M=1$ GeV by solid red and solid blue, respectively. The blue dashed line is for $\lambda_L=1.0$ and $\Delta M=5$ GeV. The orange dashed line shows the variation for a larger mass gap between the inert scalars.

\underline{$M_{H^0} < 100$ GeV}: We first focus on the region $M_{H^0} < 100$ GeV, where other heavy inert scalars fall within a narrow mass range. The first dip is around $M_{H^0}\sim 38-44$ GeV because of the resonance production of the $Z~(H^0 A^0\rightarrow Z)$ and $W^\pm~(H^0 H^\pm\rightarrow W^\pm)$ boson. The region $M_{H^0} \leq 45$ GeV is ruled out from the electroweak bound as given in equation \ref{Eq:Decay_const}. The second dip is a result of the resonance production of the SM Higgs boson through the Higgs portal annihilation ($H^0 H^0 \rightarrow h$). Furthermore, some additional shallow dips are caused by the opening of $H^0 H^0 \rightarrow W^+ W^-$ and $H^0 H^0 \rightarrow Z Z$ annihilation modes. This region has been studied at the LHC with a mono-jet signal, as discussed in reference \cite{Belyaev:2018ext}. 

\underline{$M_{H^0} > 100$ GeV:} The blue curves in figure \ref{Fig:relic} show a sharp dip around $M_{H^0} \sim 125$ GeV due to the opening up of $H^0 H^0 \rightarrow h h$ annihilation mode ($h$ is the SM Higgs boson), which enhances the annihilation cross-section and reduces the relic density. The $H^0 H^0 h h$ quartic coupling is directly proportional to the $\lambda_L$, so a small $\lambda_L$ value does not produce the observed dip in the red line. The horizontal dashed black line represents the observed total DM relic density. For $\lambda_L=0.01$ (1.0, solid blue), we observe that $H^0$ can provide the correct relic density when $M_{H^0} \sim 543~(1000)$ GeV; below these masses, the relic is underabundant. WIMP is more abundant for a small $\lambda_L$ value since a larger $\lambda_L$ increases the annihilation cross-section, decreasing relic density, as observed in the solid red and blue lines. For $|\lambda_L|\geq 0.001$, the region $100~\text{GeV} < M_{H^0}< 500~\text{GeV}$ generally remains underabundant, whereas the region $M_{H^0}> 500$ GeV may or may not be underabundant, depending on the $\lambda_L$ value~\footnote{If VLQ becomes nearly degenerate with the inert scalars, then the coannihilation of VLQ with other inert scalars and VLQ annihilation will also contribute to the relic density of $H^0$ \cite{Alves:2016bib}. This will result in an increase in the annihilation cross-section, and consequently, the relic density of $H^0$ will be smaller than that in figure \ref{Fig:relic}.}.
\paragraph{Axion relic density:} In our two-component dark matter scenario, the underabundant WIMP combines with the axion to produce the correct relic density, as observed by the Planck collaboration \cite{Aghanim:2018eyx}
\begin{equation}
\Omega_{H^0}h^2 + \Omega_{a}h^2 =\Omega_{\text{total}}h^2 = 0.120\pm 0.001 
\label{Eq:tot_relic}
\end{equation}
Axions have a vast literature, and some interesting references can be found in the following \cite{Berezhiani:1989fp,Ghosh:2023xhs} and the references therein.
The relic density of axions depends on whether the Peccei-Quinn symmetry is broken before or after the period of inflation. In scenarios where the symmetry is broken before inflation ends, axion production primarily occurs through coherent oscillations resulting from vacuum realignment. In this case, the axion relic density is given by \cite{Sikivie:2006ni, Bae:2008ue}, 
\begin{equation}
\Omega _a h^2 \simeq 0.18 \hspace{1mm} \theta ^2 \hspace{1mm} \bigg(\frac{f_a}{10^{12}~ \text{GeV}}\bigg)^{1.19} \, 
\label{Eq:axion_relic}
\end{equation}
where $\theta$ represents the initial misalignment angle of the axion, and it is of the order of unity. For $\theta=1$, the axion decay constant required to match the correct DM relic density in our WIMP-axion model is 
\begin{equation}
f_a=\Bigl( \dfrac{\Omega_{\text{total}}h^2-\Omega_{H^0}h^2}{0.18} \Bigr)^{1/1.19}~10^{12}~ \text{GeV} ~. 
\end{equation}
Equation \ref{Eq:axion_relic} suggests the axion can give the total relic density when $f_a \sim 10^{12}~ \text{GeV}$. Therefore, the overproduction of axions imposes a constrain $f_a \leq 10^{12}~ \text{GeV}$.

%%=============================================================

%======================================================================================
\section{Collider analysis}
\label{LHC-pheno}
%======================================================================================

%==================
\begin{figure}[tb!]
\centering
\subfloat[] {\label{fig_As1}\includegraphics[scale=0.45]{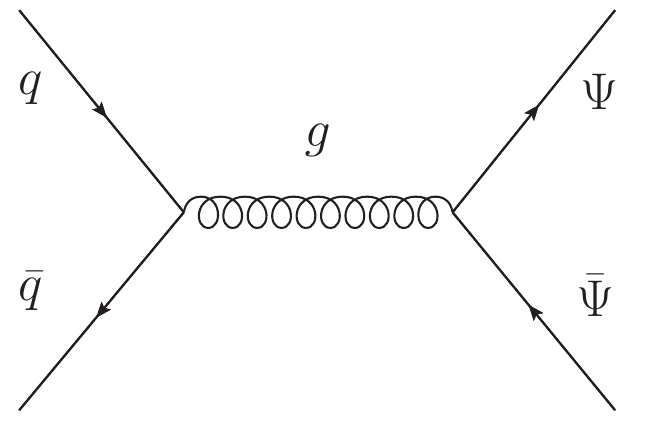}} \hspace{3pt}
\subfloat[] {\label{fig_As2}\includegraphics[scale=0.45]{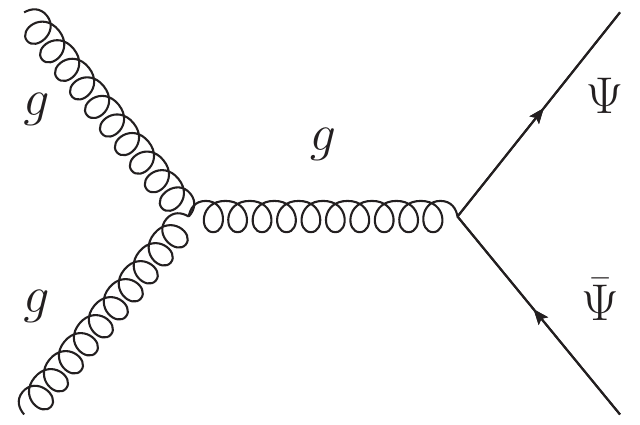}} \hspace{3pt}
\subfloat[] {\label{fig_As3}\includegraphics[scale=0.43]{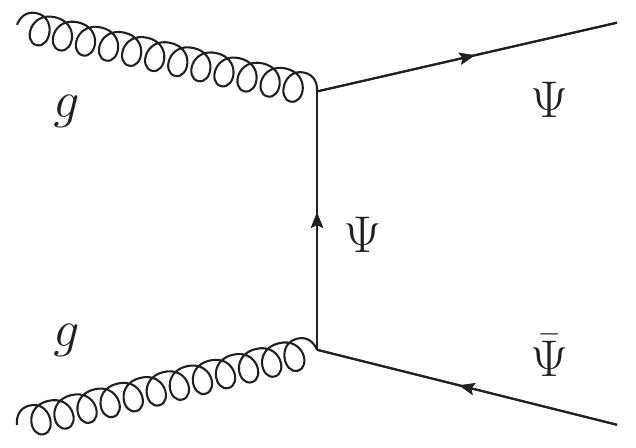}}\\ 
\subfloat[] {\label{fig_As4}\includegraphics[scale=0.40]{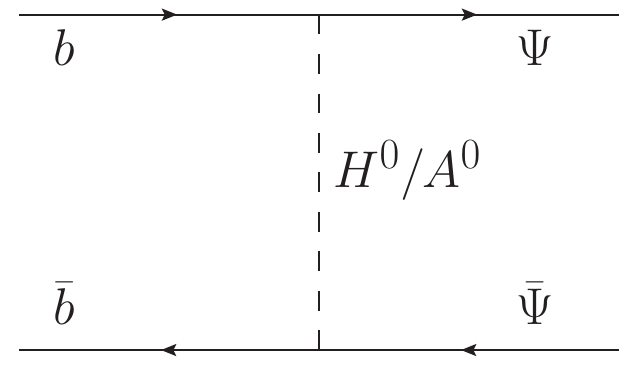}}\hspace{1cm}
\subfloat[] {\label{fig_BSM1}\includegraphics[scale=0.40]{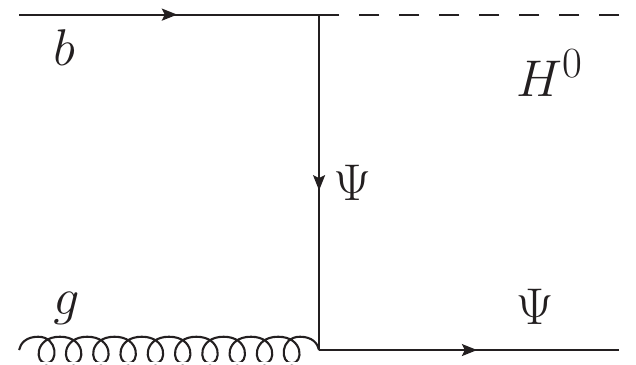}}\\ 
\subfloat[] {\label{fig_decay1}\includegraphics[scale=0.40]{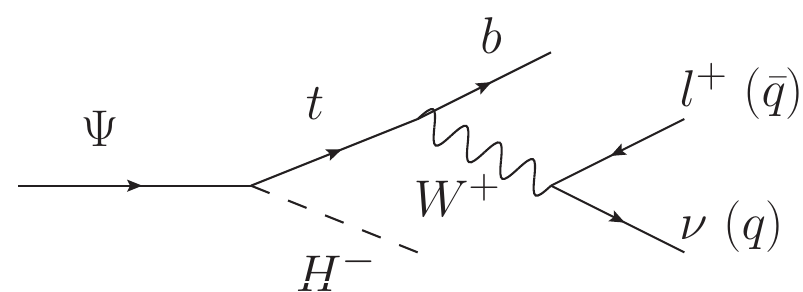}}\hspace{1cm}
\subfloat[] {\label{fig_decay2}\includegraphics[scale=0.40]{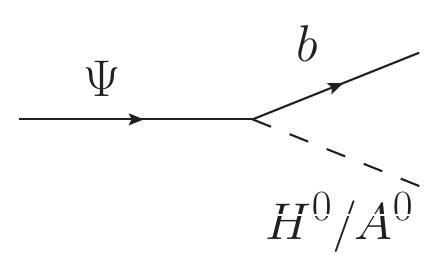}}
\caption{Feynman diagrams of $\Psi \bar{\Psi}$ pair production in the proton-proton collision at leading order partonic processes (a)-(d). Diagram (e) is the production of $\Psi$ associated with WIMP. Diagrams (f) and (g) are the possible decay modes of VLQ when it predominantly couples with the third-generation quark family of the SM.}
\label{FD_production_1}
\end{figure}
%==================

%%==================
%\begin{figure}[tb!]
%\centering
%\includegraphics[scale=0.32]{./plots/NLO_as_1.pdf}
%\includegraphics[scale=0.32]{./plots/NLO_as_2.pdf}
%\includegraphics[scale=0.32]{./plots/NLO_as_6.pdf}
%\includegraphics[scale=0.32]{./plots/NLO_as_5.pdf}\\
%\includegraphics[scale=0.32]{./plots/NLO_as_3.pdf}
%\includegraphics[scale=0.32]{./plots/NLO_as_7.pdf}
%\includegraphics[scale=0.32]{./plots/NLO_as_8.pdf}
%\includegraphics[scale=0.32]{./plots/NLO_as_9.pdf}
%\caption{}
%\label{FD_production_NLO-1_5}
%\end{figure}
%%==================

The presence of VLQ opens up interesting collider searches. VLQ can be pair-produced or produced in association with the inert scalars. The Yukawa interaction of VLQ with the SM quarks and the inert scalars is given in equation \ref{Eq:yukawa}. Considering VLQ predominantly couples with third-generation of the SM quarks, the pair production at the LHC, $p p \rightarrow \Psi \bar{\Psi}$, representative diagrams are shown in figure \ref{FD_production_1} in parts \ref{fig_As1}-\ref{fig_As4}. Quark-initiated diagram \ref{fig_As4} strongly depends on the value of the Yukawa coupling ($f$). For protons with four or five massless quark flavours, diagram \ref{fig_As4} is either absent or contributes negligibly small, even with $f \sim 1$, because of the low parton distribution function (PDF) of $b$-quarks. As a result, the first three diagrams, parts \ref{fig_As1}-\ref{fig_As3}, are only the production channels of the VLQ pair at the proton-proton collision. These processes are governed entirely by the strong interaction of the SM and are independent of any BSM couplings, though they do depend on the mass of the VLQ \footnote{In contrary, if VLQ significantly interacts with the first or both the first and second families of the SM quarks, the quark-initiated pair production processes can contribute significantly. The cross-section of these processes depends on the Yukawa coupling values and the masses of the inert scalars, making the partonic cross-section fully dependent on the model specifics.}. The production of $\Psi$ with WIMP is shown in diagram \ref{fig_BSM1}. It also has a negligibly small cross-section for $f \leq 1.0$. Therefore, to investigate this model at the LHC, the focus is solely on the pair production of VLQs. Upon their production, each VLQ decays into either a top or bottom quark, accompanied by an inert scalar, as illustrated in diagrams \ref{fig_decay1} and \ref{fig_decay2}. Depending on the decay modes of VLQ, the possible signal topologies are detailed below.
\begin{align}
& p p \rightarrow \Psi \bar{\Psi} \rightarrow (t~H^-) (\bar{t}H^+)= t \bar{t} + H^+ H^-\label{topo:1}\\
& p p \rightarrow \Psi \bar{\Psi} \rightarrow (t~H^-) (\bar{b}H^0/\bar{b} A^0)= t \bar{b} (\text{or}~ \bar{t}b ) + H^\pm H^0 / H^\pm A^0\label{topo:2}\\
& p p \rightarrow \Psi \bar{\Psi} \rightarrow (b H^0/b A^0) (\bar{b}H^0/\bar{b} A^0)= b \bar{b} + H^0 H^0 / H^0 A^0 / A^0 A^0\label{topo:3}
\end{align}
The signal we are looking for contains at least one top (or anti-top) quark that decays leptonically, so both topologies \ref{topo:1} and \ref{topo:2} will play a role. In topology \ref{topo:1}, one top quark undergoes a leptonic decay while the other decays hadronically. Conversely, topology \ref{topo:2} involves one b-jet with the top quark decaying leptonically. In both cases, we end up with a charged lepton and two b-quarks from either the top quark's decay or VLQ's. Therefore, our final state includes one isolated energetic lepton (electron or muon) and inclusive two jet events, with one of the leading two jets expected to be b-tagged. In addition, our final state has a large missing transverse momentum (MET or $\slashed{E}_T$), which comes from the neutrinos from the top decay and the nearly degenerate inert scalars. Topology \ref{topo:3} does not contribute to our desired final state, as no lepton is present. This final state is highly promising for dark matter searches at the LHC and provides a complementary search to the multijet + MET studies~\footnote{Many new physics models beyond the Standard Model are revealed using the final state involving fatjets plus MET \cite{Ghosh:2023ocz,Ghosh:2022rta}. For TeV scale exotic particle searches at the LHC, fatjets+MET provides better sensitivity than searches involving jets + MET.}. It marks an intriguing direction for exploring the degenerate spectrum of IDM and VLQ at the LHC. Additionally, we include the next-to-leading order (NLO) QCD correction of VLQ pair production processes for more accurate predictions. We then match these NLO fixed-order results with the {\sc Pythia8} parton shower, ensuring our findings are applicable over entire phase-space regions. 
%
%==================
\begin{table}[tb!]
\begin{center}
 \begin{tabular}[b]{|c|c|c|c|c|c|c|c|}
\hline\hline
\scriptsize Benchmark & \scriptsize $M_{H^0}$ & \scriptsize $M_{\Psi}$ & \scriptsize $\Delta M_{\Psi DM}$  & \scriptsize BR($\Psi\rightarrow t H^{-}$) & \scriptsize LO, $~~\mathcal{O}(\alpha_S^2)$ & \scriptsize NLO, $~~\mathcal{O}(\alpha_S^3)$   & \scriptsize K-fac\\ 
\scriptsize points    & \scriptsize (GeV) &  \scriptsize (GeV) & \scriptsize (GeV) &   & (fb) & (fb)&\\% 
\hline\hline
\scriptsize BP1 & 645  & 850 & 205 & 0.3528  & \scriptsize $89.21^{+28.6\%}_{-20.8\%}$ & \scriptsize $133.2^{+9.6\%}_{-11.0\%}$ & 1.49\\
\hline
\scriptsize BP2 & 445  & 750 & 305 & 0.40  &  \scriptsize $191.7^{+28.6\%}_{-20.8\%}$ & \scriptsize $287.2^{+9.1\%}_{-10.7\%}$ & 1.50 \\
\hline
\scriptsize BP3 & 550  & 1000 & 450 & 0.4867  & \scriptsize $31.76^{+28.7\%}_{-20.9\%}$ & \scriptsize $47.33^{+10.5\%}_{-11.5\%}$& 1.49\\
\hline
\scriptsize BP4 & 500  & 1100 & 600 & 0.494  & \scriptsize $16.67^{+28.8\%}_{-21.0\%}$ & \scriptsize $24.45^{+10.2\%}_{-11.4\%}$ & 1.47\\
\hline
\scriptsize BP5 & 700  & 1500 & 800 & 0.4964  & \scriptsize $1.79^{+29.2\%}_{-21.3\%}$ & \scriptsize $2.47^{+11.6\%}_{-12.2\%}$ & 1.38\\
\hline
\hline
 \end{tabular}
 \caption{Few representative benchmark points (BPs) satisfy all the constraints discussed in section \ref{theory}. $M_{H^0}$ and $M_{\Psi}$ are the mass of the WIMP DM and VLQ, respectively, with $\Delta M_{\Psi DM}=M_{\Psi}-M_{H^0}$. Here, we fix mass difference within inert scalars $M_{A^0}-M_{H^0}=M_{H^\pm}-M_{H^0}=5.0$ GeV. BR($\Psi\rightarrow t H^{-}$) is the branching fraction of VLQ decaying into a top quark and $H^-$, and BR($\Psi\rightarrow b H^0$) + BR($\Psi\rightarrow b A^0$) = 1-BR($\Psi\rightarrow t H^{-}$). The sixth column contains the partonic leading order cross-section of VLQ pair production, which has order $\mathcal{O}(\alpha_S^2)$. The following columns contain next-to-leading order cross-section ($\mathcal{O}(\alpha_S^3)$), and K-factor. The superscript and subscript denote the scale uncertainties (in percentage) of the total cross-section.
}\label{cross-section}
\end{center}
\end{table}
%==================

Table \ref{cross-section} lists several benchmark points that meet all the constraints outlined in section \ref{theory}. The leading order (LO) partonic cross-section, $\sigma(pp \rightarrow \Psi \bar{\Psi})$, which encompasses diagrams \ref{fig_As1}-\ref{fig_As3}, involves a coupling order $\mathcal{O}(\alpha_S^2)$, where $\alpha_S$ represents the strong coupling constant. The next-to-leading order (NLO) cross-section is of order $\mathcal{O}(\alpha_S^3)$. The LO and NLO cross-sections are presented in the table for the central choice of factorization ($\mu_F$) and renormalization ($\mu_R$) scales. The central choice is $\mu_F=\mu_R= \text{partonic centre-of-mass energy}~(\sqrt{\hat{s}})$. $\mu_F$ and $\mu_R$ are two spurious scales resulting from collinear factorization and renormalization processes. The hadronic cross-section is theoretically independent of these scales. Nonetheless, when the perturbative series is truncated at a finite order, like LO, NLO, or so on, this invariance is lost, leading to scale uncertainties in the cross-section. Such uncertainties are more prominent at lower orders.  These scales are then varied as $\mu_F= \zeta_1 \sqrt{\hat{s}}$ and $\mu_R= \zeta_2 \sqrt{\hat{s}}$, where $\zeta_1$ and $\zeta_2$ can take values from the set $\{1/2, 1, 2\}$, creating a total of nine data sets. The superscripts and subscripts in the cross-section denote the envelope of those nine scale choices. As expected, the theoretical scale uncertainties for NLO-QCD events are significantly smaller than those for LO events. The K-factor, defined as the ratio of the NLO to LO cross-section, is approximately 1.5 for a VLO mass of up to 1 TeV, but it decreases with an increase in mass. Thus, $\mathcal{O}(\alpha_S)$ corrections applied to the pair production processes increase the production rate and lessen the scale uncertainty.

%======================================================================================
\subsection{Simulation details}
%======================================================================================
Our analysis incorporates all background processes that mimic the signal, which is detailed in the following subsection. We generate each background process with an addition of two to four QCD jets using an {\sc MLM} matching scheme \cite{Mangano:2006rw, Hoeche:2005vzu}. Our model is implemented in {\sc FeynRules} \cite{Alloul:2013bka}, and we generate the UFO model file. The signal and background events are generated using {\sc MadGraph5\_aMC@NLO}~\cite{Alwall:2014hca} and pass through {\sc Pythia8}~\cite{Sjostrand:2001yu, Sjostrand:2014zea} for showering, fragmentation and hadronization. These showered events undergo detector simulation via {\sc Delphes3} \cite{deFavereau:2013fsa}. The \emph{NNPDF3.0} \cite{NNPDF:2014otw} LO and NLO PDF datasets are used to generate LO and NLO events. {\sc Anti-kT} \cite{Cacciari:2008gp} jets with a radius parameter 0.4 are formed by clustering particle-flow towers and tracks. Finally, for the multivariate analysis (MVA), we use the adaptive Boosted Decision Tree (BDT) method \cite{Roe:2004na, FREUND1995256, Freund:1997xna} within the {\sc TMVA} \cite{Hocker:2007ht} framework.

%======================================================================================
\subsection{Backgrounds}
%======================================================================================

Our signal consists of a single energetic isolated lepton (electron or muon), multijet, and a large amount of missing transverse energy. One of the two leading jets must be b-tagged, and any event with no charged lepton or more than one lepton is rejected. The background includes $t\bar{t}+\text{jets}$, $W+\text{jets}$, and single top (including $tW+\text{jets}$), among others.
\begin{itemize}
\item[I.] \underline{$t\bar{t}+\text{jets}$}\hspace{0.4cm} $t\bar{t}$ events can be categorized into two classes that will significantly contribute to our final state. In semileptonic $t\bar{t}$ events, one top-quark decay leptonically and the other decays hadronically, resulting in one prompt lepton from the decay of the $W$ boson. The other category is dileptonic $t\bar{t}$ events, where both tops decay leptonically, producing two prompt leptons.
\paragraph*{}
The semileptonic $t\bar{t}$ events contain only one charged lepton and a single neutrino from $W$ decay, which is the source of MET. This process has a lower MET due to the presence of only one neutrino. On the other hand, dileptonic $t\bar{t}$ events mimic the signal when one of the charged leptons is lost, sometimes referred to as the lost-lepton background. In dileptonic $t\bar{t}$ events, two neutrinos from $W$ decay and one lost lepton collectively give the MET. The $t\bar{t}$ background is generated with an additional two jets with MLM matching and is normalized with the higher-order QCD corrected production cross-section at the 14 TeV LHC \cite{Muselli:2015kba}.

\item[II.] \underline{Single top-quark}\hspace{0.4cm} Single top quark production at the LHC with the  $W$ boson contributes considerably as background, which is generated with two extra parton MLM matching. We consider both cases where either the top quark or $W$ boson decays leptonically or both of them decay leptonically. Therefore, in the former case, the single neutrino or, in the latter case, two neutrinos with one mistag charged lepton give the MET.
%For the semileptonic $t W+\text{jets}$ event, the top or $W$ boson decays leptonically to give a prompt charged lepton, and other decay hadronically. This background also has a lower MET because of the presence of only one neutrino. In contrast, if both the top quark and $W$ boson decay leptonically and one of the charged lepton evades detection, the contribution will be negligibly small. 
\paragraph*{} Another small contribution comes from the single top quark production associated with the $b$-quark ($t\bar{b}~ (\bar{t} b)+\text{jets}$). This background is also generated with two extra QCD radiation with MLM matching. In this case, the top decays leptonically, giving one prompt lepton and MET. The $t W$ and $t b$ backgrounds are normalized using higher-order QCD cross-section \cite{Kant:2014oha, Kidonakis:2015nna}.

\item[III.] \underline{$W+\text{jets}$} \hspace{0.4cm} The $W+\text{jets}$ can mimic our signal when the $W$ decays leptonically. This background contains only one neutrino, which gives the MET and a single prompt-charged lepton. The background is generated with up to four QCD jets using MLM matching, resulting in a final state with multijets. This background is also scaled by a higher-order QCD cross-section \cite{Balossini:2009sa}.

\item[IV.] \underline{di-boson+jets}\hspace{0.4cm} Among diboson productions, $WZ+\text{jets}$ and $WW+\text{jets}$ will mimic our final state. In the former case, $W$ decays leptonically to give rise to a lepton, and $Z$ decays invisibly ($Z\rightarrow \nu \bar{\nu}$). Those three neutrinos, two coming from $Z$ decay and one from $W$ decay, collectively produce MET. We observed that if $Z$ boson decays hadronically, it contributes much less than $Z$ decays invisibly. This is because in the hadronic decay of $Z$, the $WZ$ background yields only one neutrino with a small MET. Therefore, this background becomes significantly smaller when a large MET cut is applied. In the $WW+\text{jets}$ background, one of the $W$ decays leptonically and the other hadronically. Both backgrounds are generated with two extra jets. We scaled the di-boson productions with the higher-order QCD cross-section \cite{Campbell:2011bn}.

\item[V.] \underline{$t \bar{t} V$}\hspace{0.4cm} In our analysis, $t \bar{t}$ production with a vector boson $V~ (=W,Z)$ has a minimal contribution due to its small cross-section, but we include it. Compared to $t \bar{t} W$, $t \bar{t} Z$ has a larger contribution when one of the top quarks decays leptonically, and the $Z$ boson decays invisibly.

\end{itemize}

%======================================================================================
\subsection{Event reconstruction}
\label{reco}
%======================================================================================

The events in our analysis can be selected using triggers with a large MET and the presence of an isolated electron or muon. MET is defined as the negative vector sum of the transverse momentum of all particle flow candidates \cite{CMS:2017yfk}. We chose MET to be greater than 80 GeV. The lepton must have transverse momenta $P_T(l)>30$ GeV and must also fall within the detector's acceptance, having a pseudo-rapidity $|\eta|$ less than 2.4. Additionally, if electrons and positrons fall within the range $1.44 < |\eta| < 1.57$, we reject the event since this range is the transitional area between the barrel and endcap sections of the CMS electromagnetic calorimeter. The lepton should have passed the isolation criteria. 
\paragraph{Lepton Isolation:} Isolation of leptons $(l=\mu,e)$ is implemented via the following variable,
\begin{equation}
I_{l} =\sum_{i(\neq l)\subset \text{tracks}}^{P_T(i)>P_T^{\text{min}}} P_T(i)~.
\end{equation}
The sum runs over all tracks above transverse momentum $P_T^{\text{min}}=0.5$ GeV, excluding the lepton of interest, lying within a cone of radius $\Delta R= \sqrt{(\Delta \eta)^2+(\Delta \phi)^2}=0.3$ centred on the lepton. Muon isolation requires $I_\mu \leq 0.1\times P_T(\mu)$, and electron isolation requires $I_e < 5$ GeV.
\paragraph{Lepton veto:} In case the event contains more than one lepton, we use the lepton veto. We discard such events if any additional lepton aside from the leading lepton has transverse momentum $P_T (l) > 5$ GeV and $|\eta| < 2.4$. 

Di-leptonic $t\bar{t}+\text{jets}$ background events can contribute to our desired final state when one lepton passes all the criteria defined above, but the second lepton passes the loose selection (lepton veto) or the lepton isolation criteria. Furthermore, the final state contains at least two jets. Jets are formed by clustering particle flow candidates with a radius parameter 0.4, utilizing the {\sc Anti-kT} algorithm. Each leading and subleading jets are required $P_T(j_i)~(i=1,2) > 30$ GeV and $|\eta| < 2.4$. The events in which the leading two jets or the isolated lepton overlap with the missing transverse momentum within a cone of radius $\Delta R < 0.4$ are discarded. All those cuts mentioned above will be referred to as a preselection cut.

%======================================================================================
\subsection{Preselection and high-level variables}
\label{selection}
%======================================================================================

%==================
\begin{figure}[tbh!]
\centering
\subfloat[] {\label{pre:MTlmet} \includegraphics[width=0.48\textwidth]{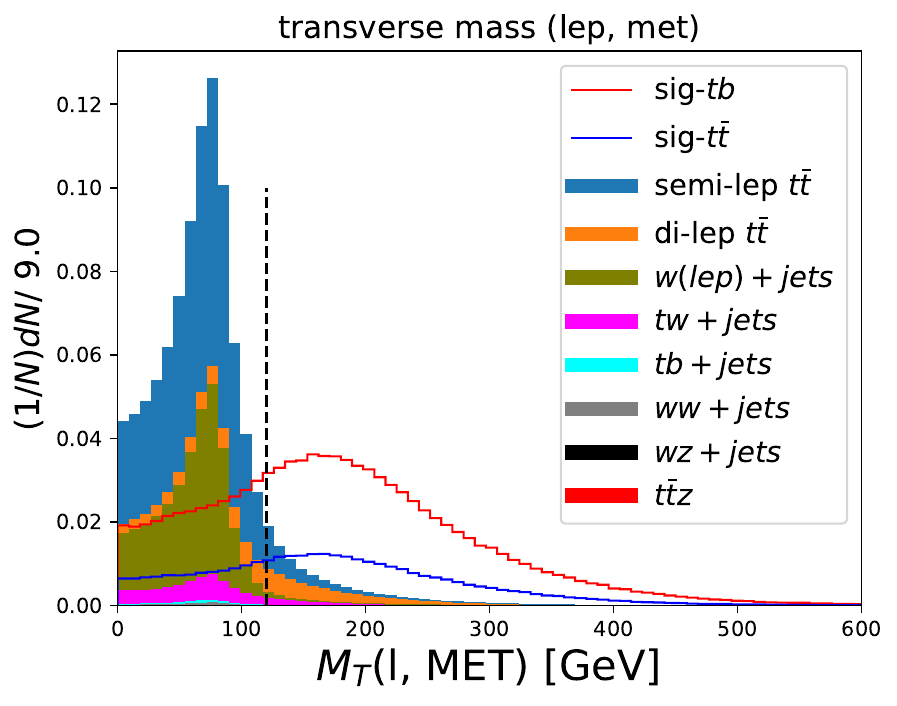}} %\hspace{-0.1cm}
\subfloat[] {\label{pre:met} \includegraphics[width=0.48\textwidth]{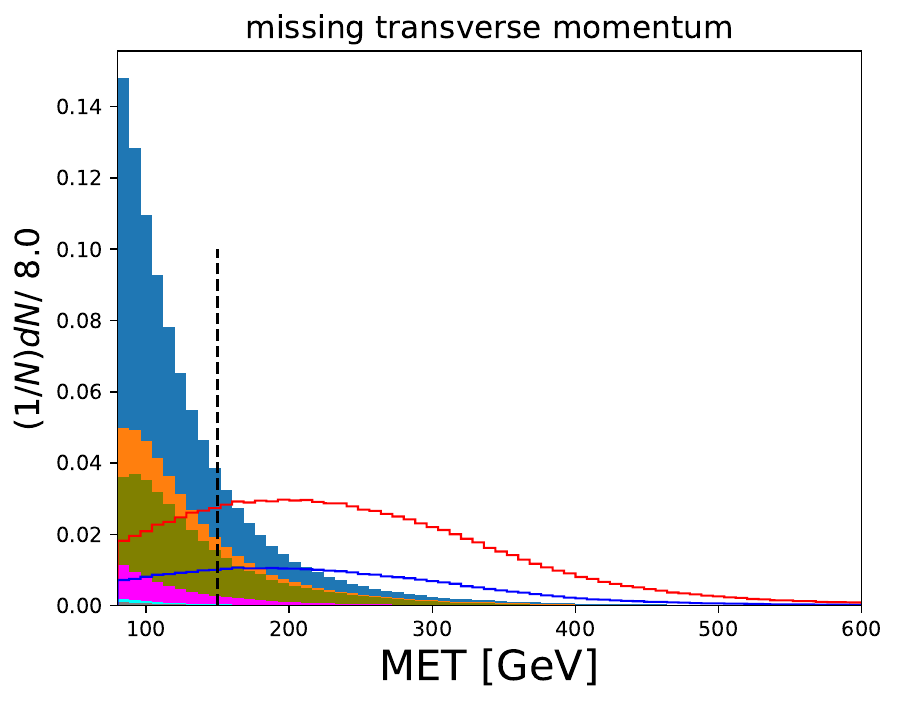}}
\caption{The normalized stacked histograms of the signal (BP2) and all background processes are shown. On the left, the distribution of the transverse mass of the lepton is presented, while on the right, the distribution of the total missing transverse momentum after preselection cuts is displayed. The signals sig-$t\bar{t}$ and sig-$tb$ correspond to topologies given in equations \ref{topo:1} and \ref{topo:2}, respectively. The decay branching ratio of $\Psi$ is detailed in table \ref{cross-section}.}
\label{fig:sig_bg_1}
\end{figure}
%==================

After preselection cuts, the distribution of the transverse mass of the lepton is shown in the left panel of figure \ref{fig:sig_bg_1}. In this stacked histogram, the contribution of individual background processes is filled with different colors. The signal is presented for BP2. The signals sig-$t\bar{t}$ and sig-$tb$ correspond to topologies in equations \ref{topo:1} and \ref{topo:2}, respectively. The solid red line represents the total signal distribution. The contribution of sig-$tb$ lies between the blue and red lines, while the solid blue line represents the contribution of sig-$t\bar{t}$. The transverse mass of the lepton is defined as \cite{Cheng:2008hk},
\begin{equation}
M_T(l,\text{MET})=\sqrt{2 P_T(l) \slashed{E}_T (1-\cos\Delta \Phi)}
\label{eq:MTl}
\end{equation}
where $P_T(l)$ denotes the lepton's transverse momentum, and  $\Delta \Phi$ represents the azimuthal separation between the lepton's direction and the missing transverse momentum (MET or $\slashed{E}_T$). After preselection cuts, the main backgrounds that mimic our signal are the semileptonically decaying $t\bar{t}+$ jets (filled blue) and the leptonically decaying $W+$ jets (filled olive). In contrast, the dileptonic $t\bar{t}+$ jets (filled orange) background is smaller. The vertical black dashed line corresponds to a lepton's  transverse mass of 120 GeV. Applying this cut significantly reduces the semileptonic $t\bar{t}$ and $W+$ jets backgrounds, making the dileptonic $t\bar{t}+$ jets background the major one, followed by the semileptonic $t\bar{t}$. 

\begin{figure}[t]
\centering
\subfloat[] {\label{pt_j1} \includegraphics[width=0.33\textwidth]{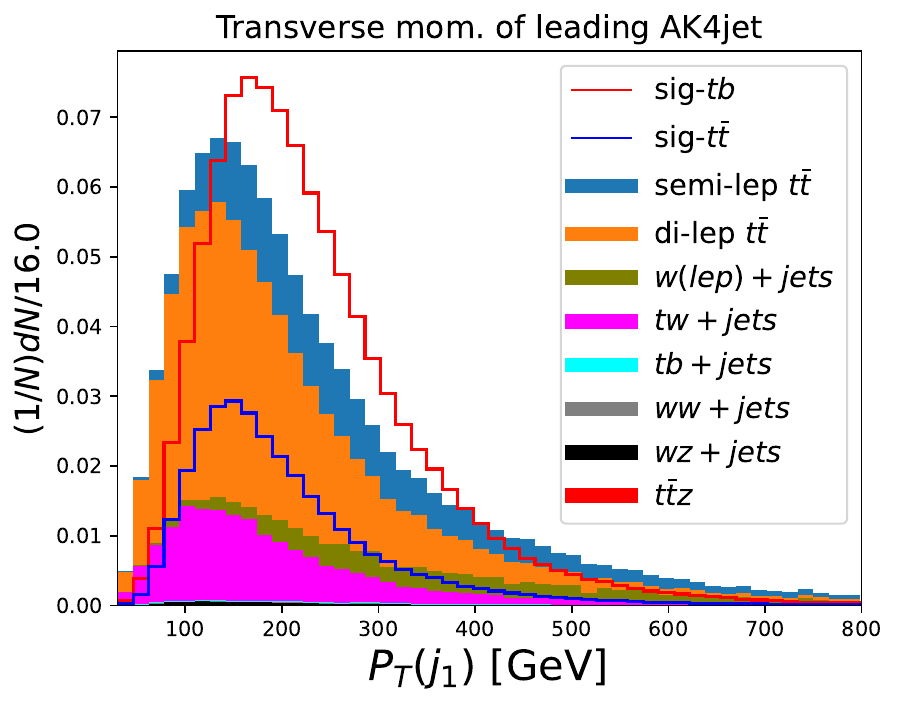}} %\hspace{-0.1cm}
\subfloat[] {\label{pt_j2} \includegraphics[width=0.33\textwidth]{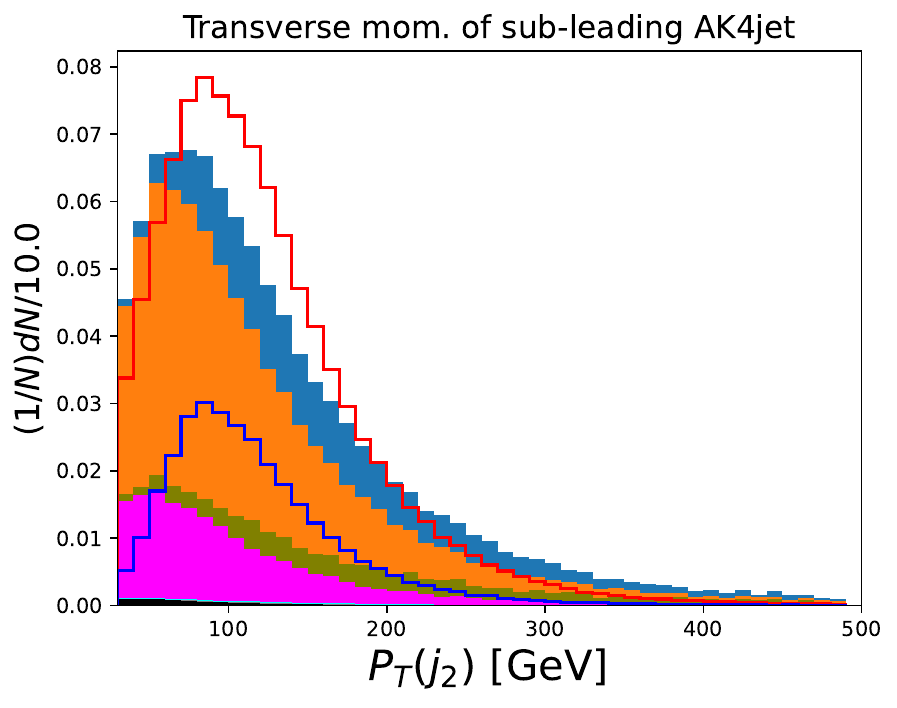}}
\subfloat[] {\label{pt_lep} \includegraphics[width=0.33\textwidth]{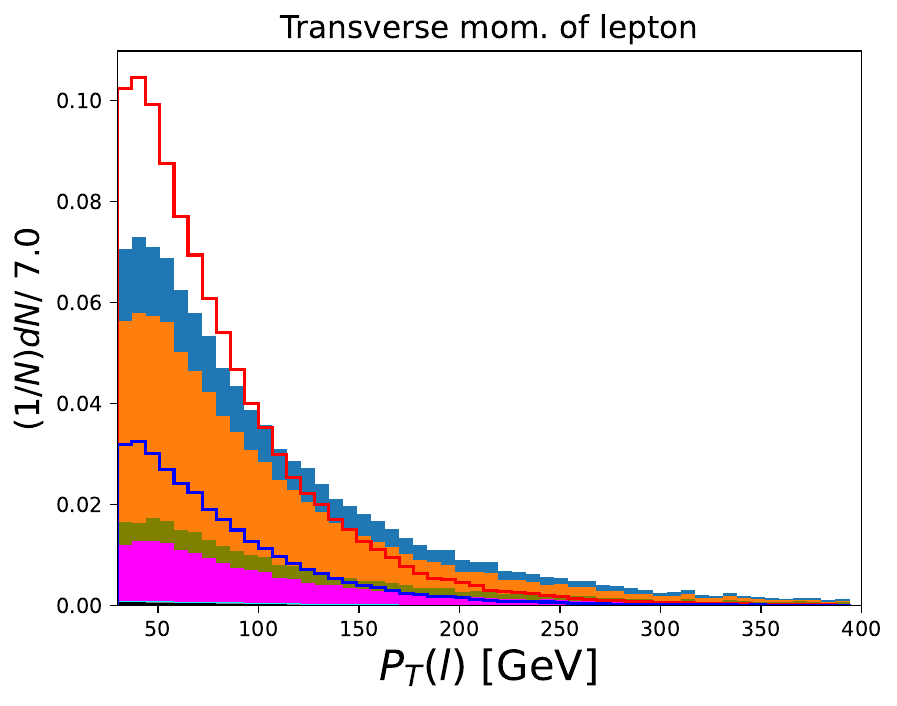}}\\
\subfloat[] {\label{met} \includegraphics[width=0.33\textwidth]{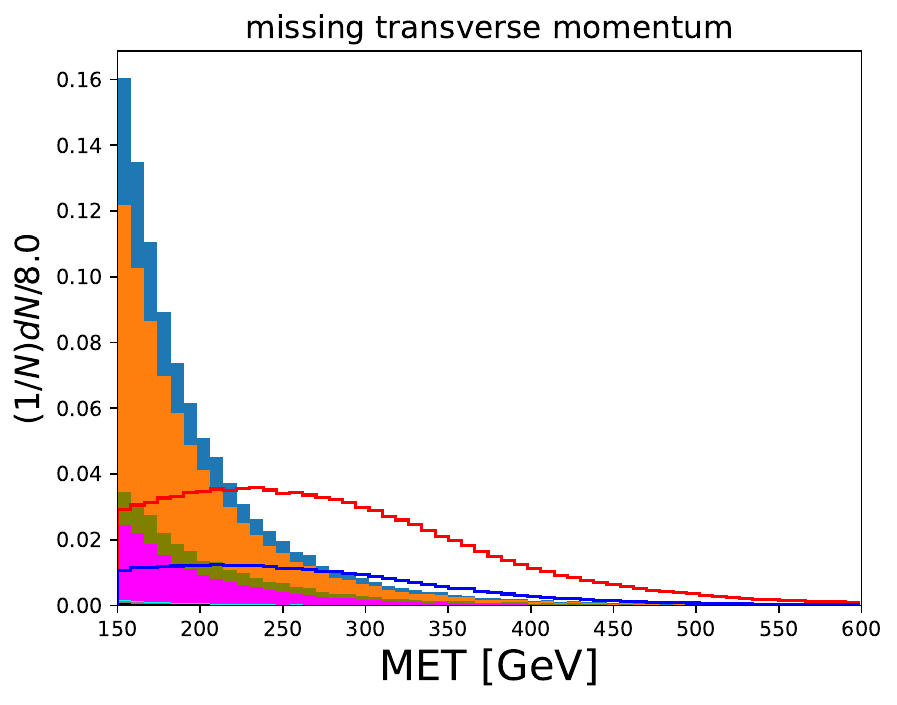}} 
\subfloat[] {\label{MT2} \includegraphics[width=0.33\textwidth]{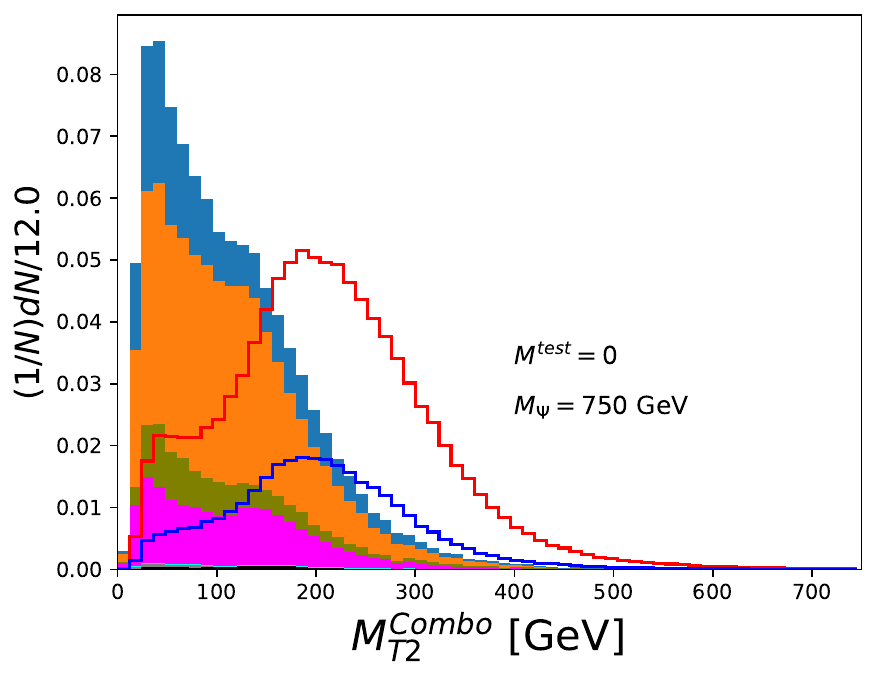}}
\subfloat[] {\label{shat} \includegraphics[width=0.33\textwidth]{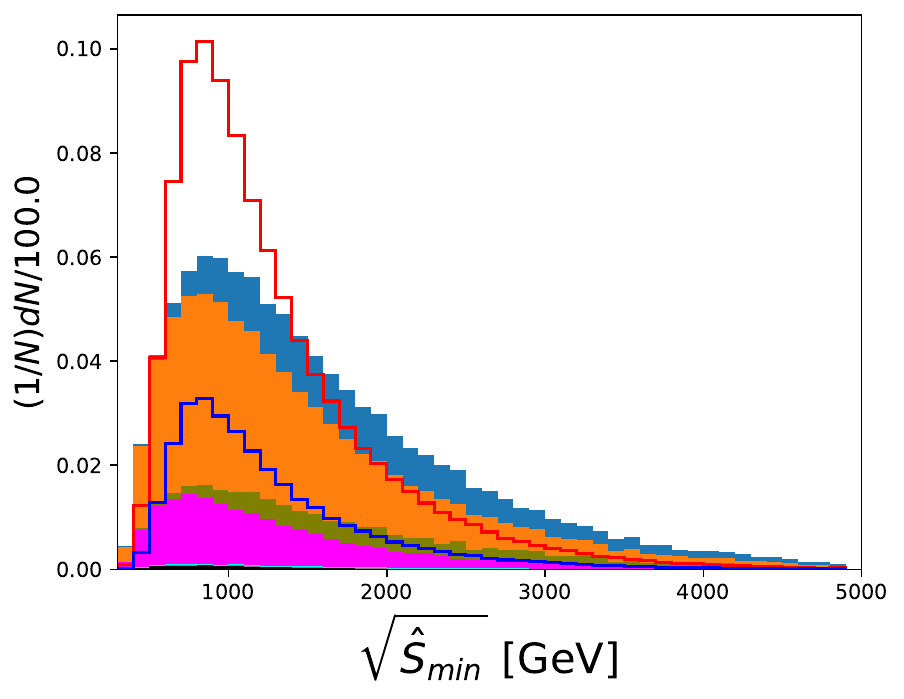}}\\
\subfloat[] {\label{MT_bmet} \includegraphics[width=0.33\textwidth]{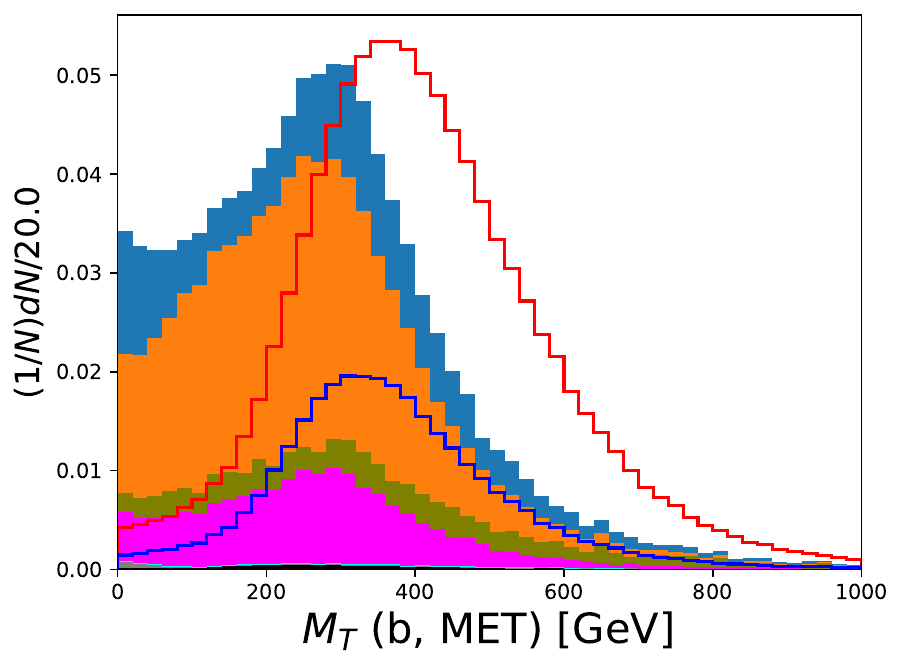}} 
\subfloat[] {\label{MT_lmet} \includegraphics[width=0.33\textwidth]{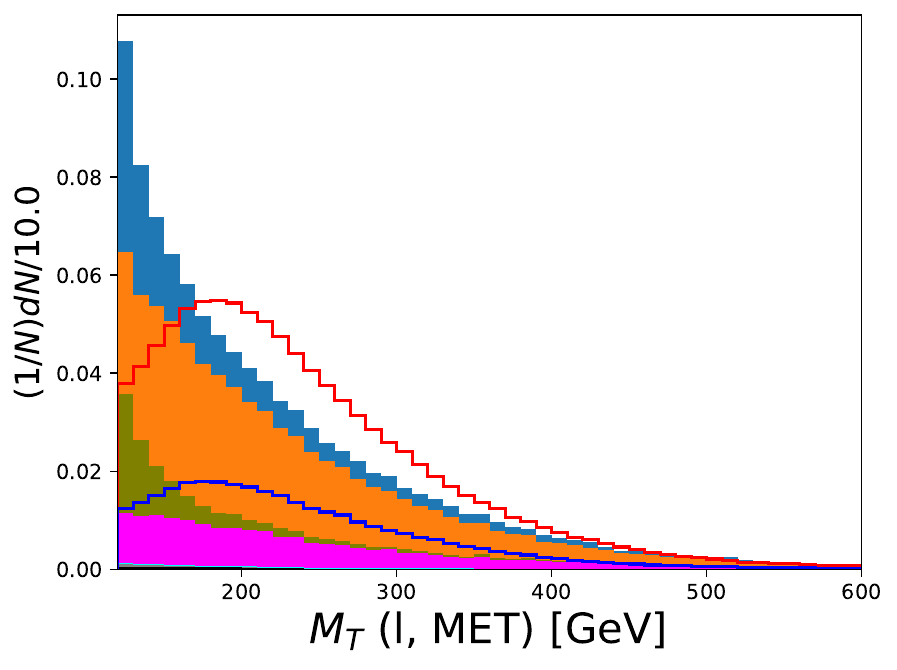}}
\subfloat[] {\label{DR_metl} \includegraphics[width=0.33\textwidth]{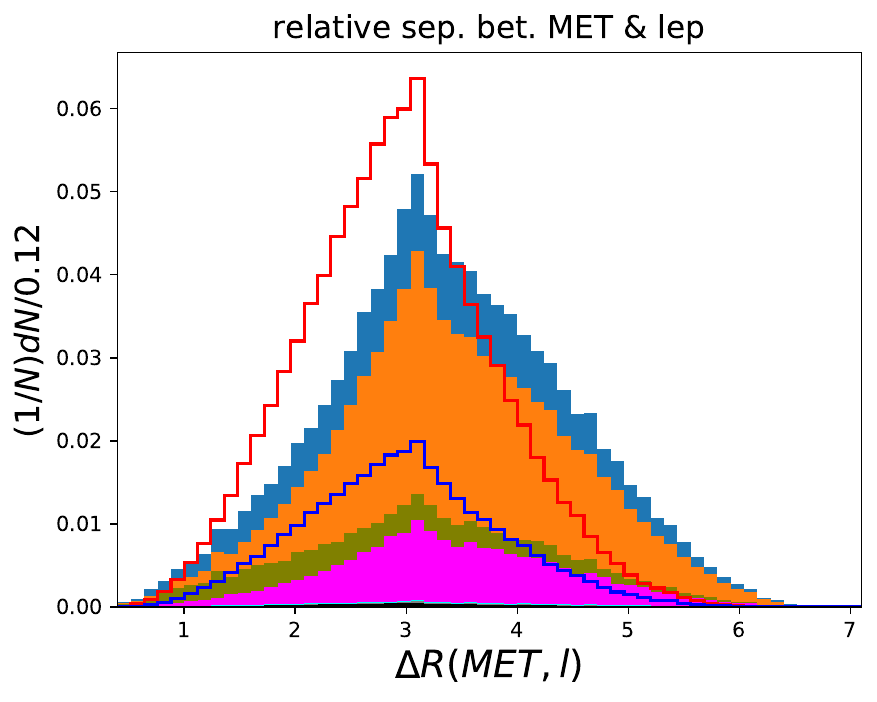}}\\
\subfloat[] {\label{DR_bl} \includegraphics[width=0.33\textwidth]{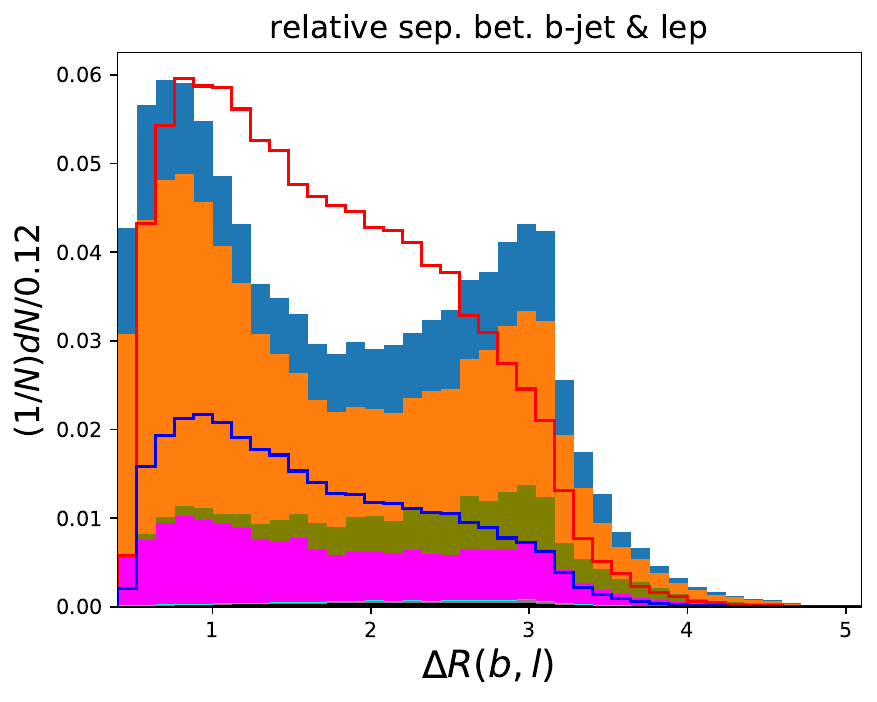}} 
\subfloat[] {\label{DPhi_metj1} \includegraphics[width=0.33\textwidth]{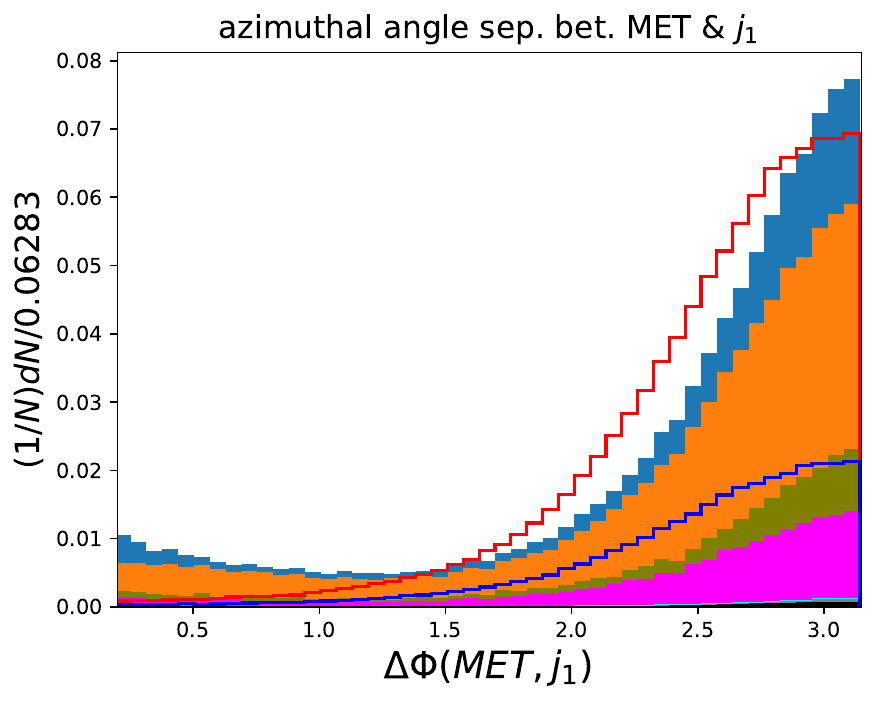}}
\subfloat[] {\label{Njets} \includegraphics[width=0.33\textwidth]{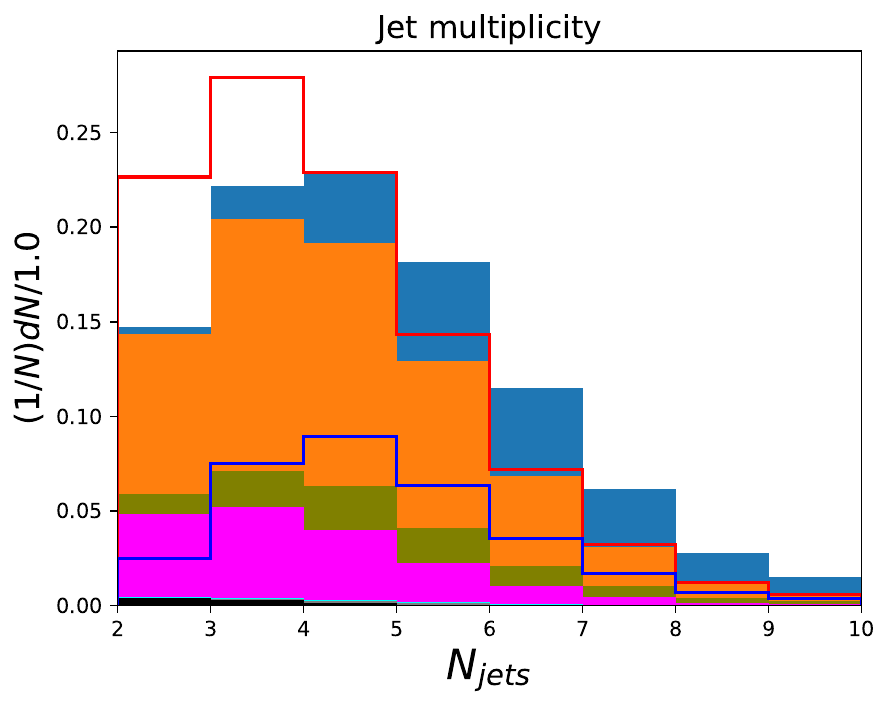}}
\caption{The normalized stacked histograms for the signal (BP2) and background processes are shown after applying preselection cuts along with the lepton transverse momentum $> 120$ GeV and missing transverse momentum $> 150$ GeV.}
\label{fig:sig_bg_MVA}
\end{figure}

The right panel of figure \ref{fig:sig_bg_1} shows the stacked histogram of the missing transverse momentum for the signal (BP2) and background. The vertical black dashed line in this plot indicates a MET of 150 GeV. All the background has a smaller MET, while the signal has a larger MET. Hence, using a higher MET cut increases the signal-to-background ratio. As a result, we apply $\text{MET} > 150$ GeV and $M_T(l,\text{MET}) > 120$ GeV along with the preselection cuts for the actual Multivariate analysis (MVA) in the next section. The signal and background distribution of different kinematic variables after applying those cuts are presented in figure \ref{fig:sig_bg_MVA}. 

Semileptonic $t\bar{t}$ and $W+$ jets are the reducible backgrounds, and by applying cuts on lepton's transverse mass and MET, we reduce these two backgrounds significantly. However, dileptonic $t\bar{t}$ remains an irreducible background with signal-like distribution. In signal, degenerate inert scalars and neutrino (from the top quark decay) give a large MET. In the dileptonic $t\bar{t}$ background, two neutrinos and a mistagged lepton contribute to a large MET. Furthermore, multiple sources of missing particles do not lead to any peak of lepton's transverse mass around the $W$ boson mass. As a result, we find a significant dileptonic $t\bar{t}$ background left even after a large MET and transverse mass cut. We have constructed the following variable \cite{Lester:1999tx, Barr:2011xt, Konar:2009wn}.
\begin{equation}
M_{T2} (\vec{P}_T^{~1},\vec{P}_T^{~2},\vec{\slashed{E}}_T ) = \min_{\vec{q}_T^{~1}+\vec{q}_T^{~2}=\vec{\slashed{E}}_T}   [~ \max \{~  M_{T}(\vec{P}_T^{~1},\vec{q}_T^{~1}) ,~   M_{T}(\vec{P}_T^{~2},\vec{q}_T^{~2}) ~\} ]
\label{Eq:MT2_1}
\end{equation}
where $\vec{q}_T^{~i}$ is the transverse momentum of an invisible particle, assumed to be massless, contributing to the missing transverse momentum $\vec{\slashed{E}}_T$; $\vec{P}_T^{~i}$ is the transverse momentum of a visible particle. Transverse mass is defined below,
\begin{equation}
M_{T}(\vec{P}_T^{~i},\vec{\slashed{E}}_T ) =\sqrt{2 |\vec{P}_T^{~i}| ~|\vec{\slashed{E}}_T | ~[1-\cos\Delta \Phi(\vec{P}_T^{~i},~\vec{\slashed{E}}_T )]}~.
\label{Eq:MT}
\end{equation}
In order to reduce the leading dileptonic $t\bar{t}$ background, we propose the variable $M_{T2}^{Combo}$, which is defined below. 
\begin{equation}
M_{T2}^{Combo}= \text{min} [~M_{T2} (j_1+l,~j_2,~\vec{\slashed{E}}_T),~M_{T2} (j_1,~j_2+l,~\vec{\slashed{E}}_T )  ] ~.
\label{Eq:MT2_2}
\end{equation}
The two leading jets are the $j_1$ and $j_2$, with one being $b$-tagged. The reconstructed charged lepton is denoted by $l$. The endpoint of the variable $M_{T2}^{Combo}$ is expected to be around the top quark mass $(m_t)$ for the dileptonic $t\bar{t}$ events. As a result, the backgrounds have a smaller value of $M_{T2}^{Combo}$, whereas the signal has a larger value because of the presence of many undetected particles. Therefore, this variable plays a crucial role in distinguishing the signal from the background in multivariate analysis. The distribution of the $M_{T2}^{Combo}$ variable is depicted in figure \ref{MT2}, where there is a clear separation between the signal and background; most background events have a smaller $M_{T2}^{Combo}$ value, while the signal shows a significantly larger value. 

Transverse mass distributions of the leading two jets and the lepton are shown in figures \ref{pt_j1}, \ref{pt_j2}, and \ref{pt_lep}, respectively. We can see that the shapes of those normalized distributions of signal and background are nearly the same, but the jets in the signal are slightly harder than those in the background. The MET distribution is presented in figure \ref{met}, where we see that at larger MET, the background falls sharply. In contrast, because of many missing particles in the signal, MET is large and remains nearly flat up to a certain point. Therefore, MET is a good discriminator for distinguishing between the signal and background. Another inclusive variable, $\sqrt{\hat{S}_{min}}$, is plotted in figure \ref{shat}, defined as $\sqrt{\hat{S}_{min}} = \sqrt{(\sum_j E_j)^2 - (\sum_j P_{z,j})^2} + \slashed{E}_T $ \cite{Konar:2008ei}.
%
%\begin{equation}
%\sqrt{\hat{S}_{min}} = \sqrt{(\sum_j E_j)^2 - (\sum_j P_{z,j})^2} + \slashed{E}_T 
%\label{Eq.shat}
%\end{equation}
%
Here, the sum includes all visible objects, such as jets and leptons; $E_j$ and $P_{z,j}$ represent the energy and $Z$-component of the particle's momentum.

%==================
%==================

The separation of the lepton from the MET and the $b$-jet in the $\eta-\phi$ plane is illustrated in figures \ref{DR_metl} and \ref{DR_bl}. The azimuthal separation of the leading jet from the MET and the jet multiplicity are depicted in figures \ref{DPhi_metj1} and \ref{Njets}, respectively. The distribution of the transverse mass of $b$-jet and the lepton are shown in figures \ref{MT_bmet} and \ref{MT_lmet}. The signal includes two $b$-quarks; the one with higher transverse momentum is tagged as a $b$-jet accompanied by several missing particles, making the azimuthal separation between the $b$-jet and MET arbitrary. Additionally, MET in the signal is large. This results in a large $b$-jet's transverse mass (equation \ref{Eq:MT}) in the signal compared to the background, establishing it as a good discriminator. 

%======================================================================================
\subsection{Multivariate analysis}
%======================================================================================

We select several properties of the reconstructed lepton ($l = e, \mu $), jets, and missing transverse momentum (MET or $\slashed{E}_T$) as input variables to the BDT. Specifically, we consider their transverse momentum ($P_T(i)$ for $i \in [j_1, j_2, l]$), azimuthal separation and distances in the $\eta-\phi$ plane ($\Delta\Phi(i, \slashed{E}_T)$ and $\Delta R(i, j)$ with $i\neq j \in [j_1, j_2, l, \text{b-jet}]$), where $\text{b-jet}$ is either the leading or subleading jet. Additionally, we include the invariant mass of the lepton and $\text{b-jet}$ pair $m(l,b)$, the transverse mass of the lepton ($M_T(l, \slashed{E}_T)$) and $b$-jet ($M_T(b, \slashed{E}_T)$), and the number of jets ($N_{\text{jets}}$). Moreover, we include the variable $M_{T2}^{Combo}$ as defined in equation \ref{Eq:MT2_2}, and a global inclusive variable $\sqrt{\hat{s}_{\text{min}}}$ that determines the mass scale of new physics. % ($N_{\text{jets}}$​)  $M_{T2}$​

The full set of BDT input variables is thus:
\begin{equation}
\begin{aligned}
\Biggl\{ & N_{\text{jets}},~ \slashed{E}_T,~ P_T(j_1),~ P_T(j_2),~ P_T(l),~ \Delta\Phi(j_1, \slashed{E}_T),~ \Delta\Phi(j_2, \slashed{E}_T),~ \Delta\Phi(l, \slashed{E}_T),~ \Delta R(j_1, \slashed{E}_T),\\ 
& \Delta R(j_2, \slashed{E}_T),~ \Delta R(l, \slashed{E}_T),~ \Delta R(b,l),~ m(l,b),~ M_T(l, \slashed{E}_T),~ M_T(b, \slashed{E}_T),~ \sqrt{\hat{s}_{\text{min}}},~ M_{T2}^{Combo} \Biggr\}
 \end{aligned}
\label{Eq.BDT inputs}
\end{equation}
The distribution of some of those variables after applying preselection cuts along with a transverse mass of lepton $> 120$ GeV and MET $> 150$ GeV are shown in figure \ref{fig:sig_bg_MVA}. Note that we do not use stringent preselection criteria for the input data sets in multivariate analysis. Thus, MVA has enough scope to find an optimal nonlinear boundary by applying cuts to appropriate variables and providing better statistical significance for the signal over the background. 

Multiple processes contribute to the signal and background; therefore, we take the weighted sum of those processes and make a combined signal and background class. We use the adaptive Boosted Decision Tree (BDT) algorithm in our analysis, and each signal and background event sample is split randomly for training and testing purposes. The variables mentioned in equation \ref{Eq.BDT inputs} are not all linearly independent, and some are particularly important for distinguishing between signal and background classes. Therefore, we select variables that are not highly correlated (or highly anti-correlated) in both signal and background classes and have significant importance in discriminating the signal from the background. By definition, the invariant mass of $b$-jet and lepton $m(l,b)$ depends on the azimuthal separation between $b$-jet and lepton, $\Delta\Phi(b,l)$. This relationship reveals a significant correlation between $m(l,b)$ and $\Delta R(b,l)$ in both signal and background classes. In our final selection of variables, $\Delta R(b,l)$ is preferred over $m(l,b)$ due to its greater relative importance in signal-background separation. The relative importance is obtained from the {\sc TMVA} package. Another example is that $P_T(l)$ is highly correlated with $M_{T2}^{Combo}$ in both signal and background classes, and we choose to keep $M_{T2}^{Combo}$ because of its greater relative importance compared to $P_T(l)$. 

\begin{figure}[tb!]
\centering
  \subfloat {\label{correlation_S}\includegraphics[width=0.5\textwidth]{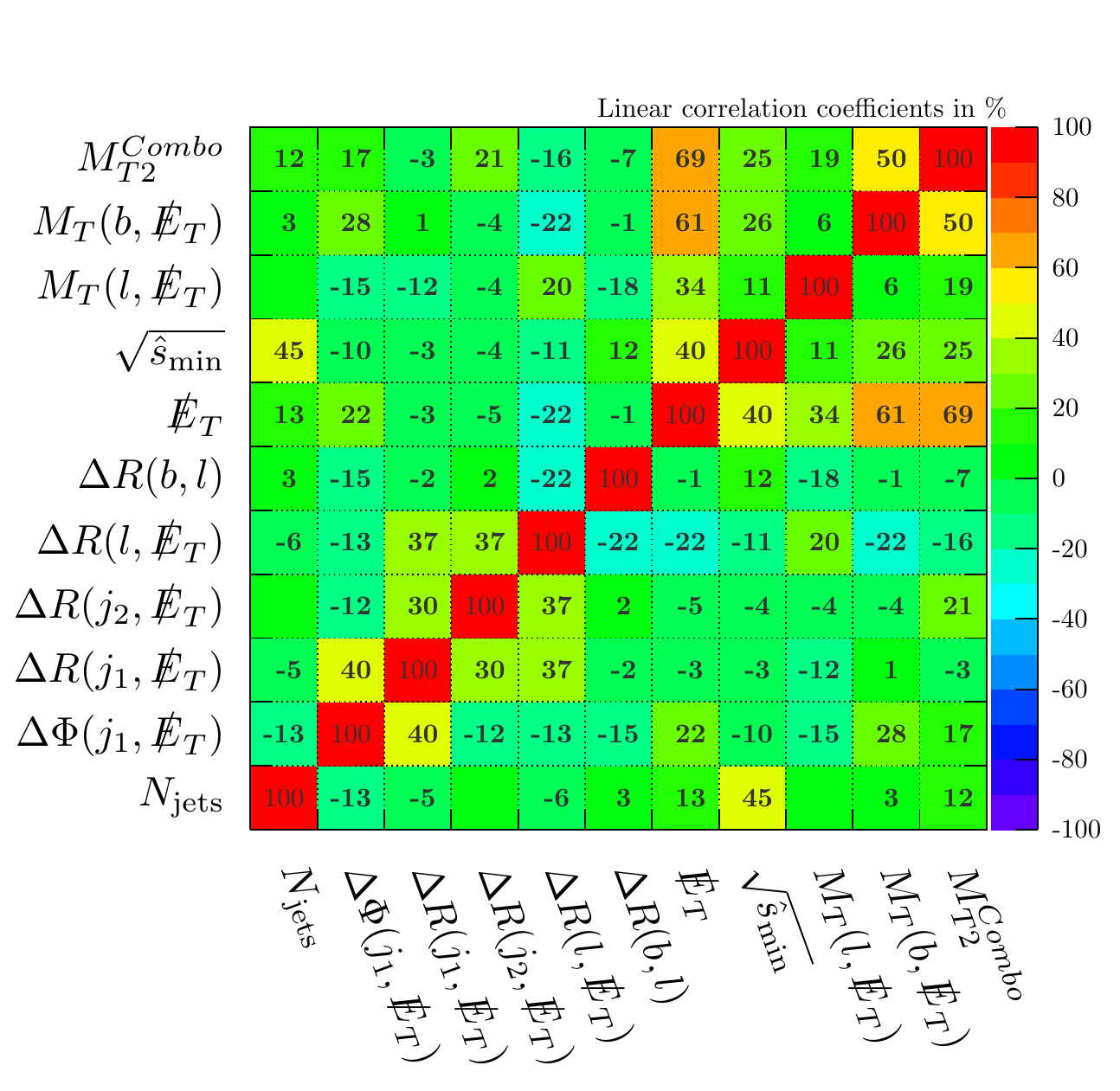}} %\hspace{-0.2cm}
  \subfloat {\label{correlation_B}\includegraphics[width=0.5\textwidth]{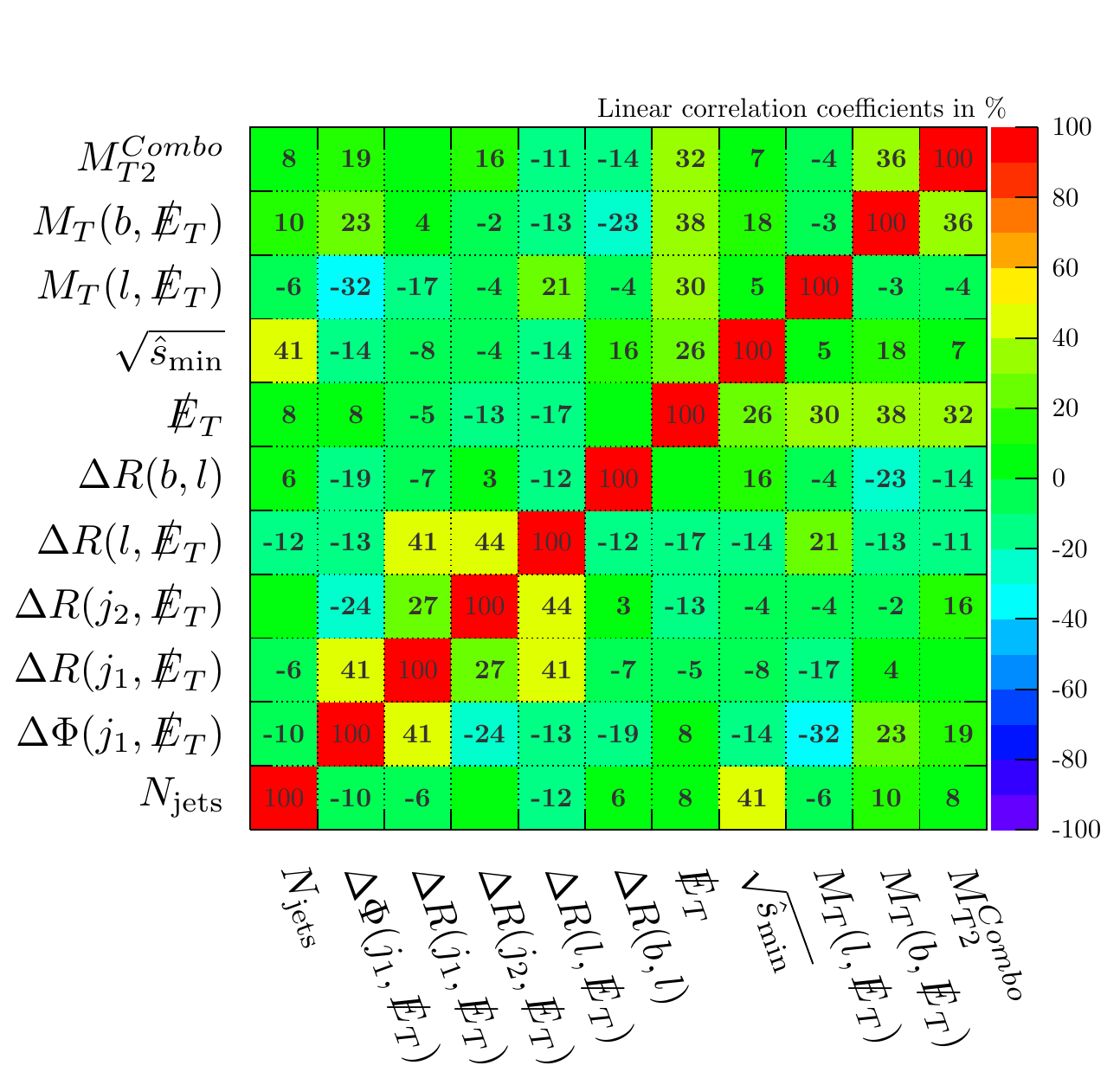}}
  \caption{Linear correlation coefficients (in percentages) between various kinematic variables for the signal (left, for BP2) and corresponding background (right) are displayed. Positive coefficients indicate correlation between variables, while negative coefficients indicate anti-correlation.}\label{correlation}
\end{figure}

The final variables are used in our multivariate analysis, and their linear correlation among themselves in signal (left panel) and background (right panel) are shown in figure \ref{correlation}. Positive coefficients indicate correlation, while negative coefficients indicate anti-correlation. If a coefficient is not mentioned, the correlation or anti-correlation between the variables is less than $1\%$. $M_{T2}^{Combo}$ and transverse mass of $b$-jet have a large correlation with the MET for the signal class but not in the background class; therefore, we keep all of them in our analysis.

%==================
%==================

%=====================
\begin{table}[tb!]
\centering
\setlength\tabcolsep{1.2pt} % default value: 6pt
 \begin{tabular}[b]{|c|c|c|c|c|c|c|c|c|c|c|c|c|}
\hline
\scriptsize & $\slashed{E}_T$ & \scriptsize $M_{T2}^{Combo}$ & \scriptsize $M_T (b,\slashed{E}_T)$ & \scriptsize $\Delta R (l,\slashed{E}_T)$ & \scriptsize $\Delta R (j_1,\slashed{E}_T)$  & \scriptsize $\Delta \Phi(j_1,\slashed{E}_T)$ & \scriptsize $\sqrt{\hat{s}_{\text{min}}}$ & \scriptsize $\Delta R (b,l)$ & \scriptsize $M_T (l,\slashed{E}_T)$ & \scriptsize $\Delta R (j_2,\slashed{E}_T)$ & \scriptsize $N_{\text{jets}}$ \\ % 
\hline\hline
\scriptsize BP1 & \hspace{0.08cm}  11.32  & \hspace{0.08cm} 13.32 & 5.761 & 8.544 & 5.486 & 4.766 & 6.994 & 5.814 & 3.931 & 3.982 & 3.06 \\
\hline
\scriptsize BP2 & \hspace{0.08cm}  26.86  & \hspace{0.08cm} 25.21 & 16.40 & 6.412 & 5.897 & 5.427 & 5.134 & 5.104 & 3.854 & 3.666 & 2.811 \\
\hline
\scriptsize BP3 & \hspace{0.08cm}  48.62  & \hspace{0.08cm} 39.76 & 32.97 & 9.47 & 9.092 & 5.538 & 5.419 & 3.915 & 11.43 &  6.062 & 1.838 \\
\hline
\scriptsize BP4 & \hspace{0.08cm}  59.62  & \hspace{0.08cm} 46.84 & 42.64 & 10.49 & 10.48 & 5.321 & 8.801 & 3.665 & 18.48 &  6.595 & 1.409 \\
\hline
\scriptsize BP5 & \hspace{0.08cm}  70.16  & \hspace{0.08cm} 54.57 & 52.99 & 13.99 & 14.20 & 5.429 & 15.87 & 3.379 & 28.32 &  8.949 & 1.508 \\
\hline
 \end{tabular} 
\caption{Variable ranking by relative importance, independent of the method used, among the observables.}\label{relative_imp}
\end{table}
%=====================
%==================
\begin{figure}[tb!]
\centering
\subfloat[] {\label{met_com} \includegraphics[width=0.32\textwidth]{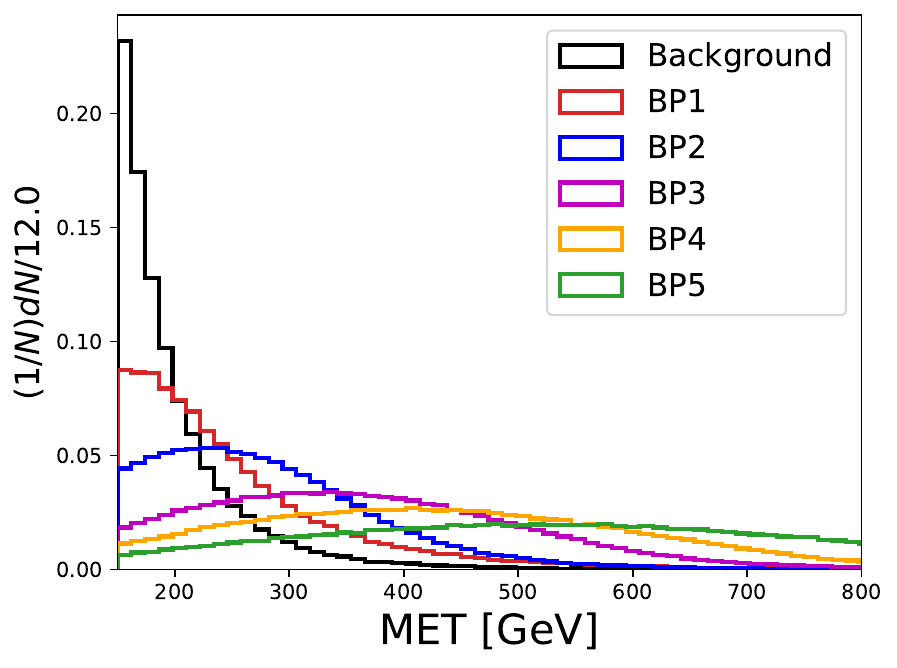}} %\hspace{-0.1cm}
\subfloat[] {\label{MT2_com} \includegraphics[width=0.32\textwidth]{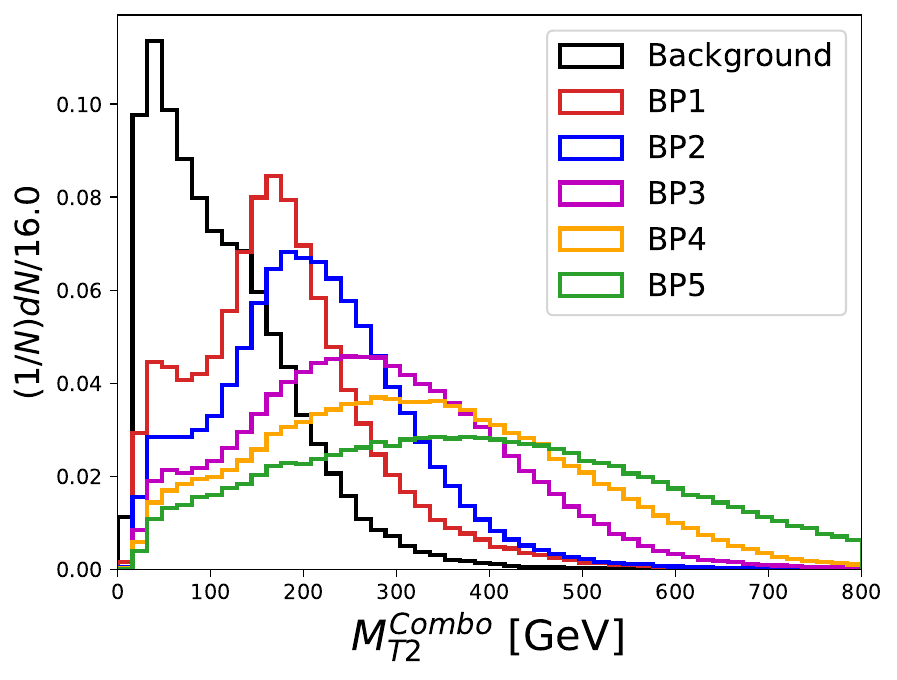}} 
\subfloat[] {\label{MT_com} \includegraphics[width=0.32\textwidth]{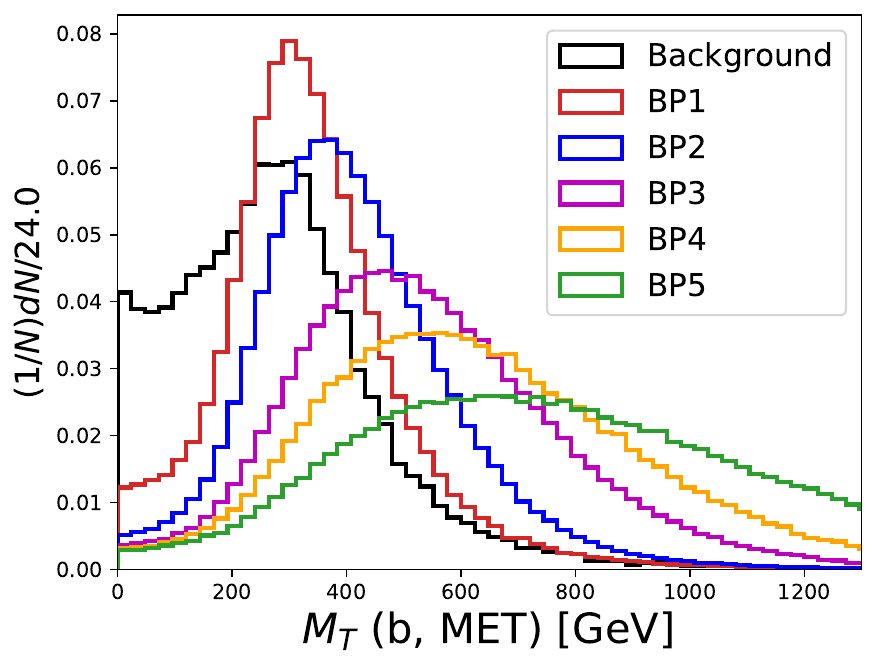}} 

\caption{The normalized histograms of the combined signal for BP1-BP5 and combined background (solid black) are shown: (a) missing transverse momentum, (b) variable $M_{T2}^{Combo}$ as defined in equation \ref{Eq:MT2_2}, and (c) transverse mass of $b$-jet. Combined indicates the weighted sum of all the processes.}
\label{fig:combined_plots}
\end{figure}
%==================

The relative importance of the variables is summarized in table \ref{relative_imp}. In our analysis, we separately optimize for all benchmark points. For BP1, all the variables are almost equally important during MVA. As the mass difference between the vector-like quark ($\Psi$) and DM ($\Delta M_{\Psi DM}$) increases towards BP5, the importance of missing transverse momentum, $M_{T2}^{Combo}$, and the transverse mass of the $b$-jet and lepton increases significantly compared to the other variables in the same row. Figure \ref{fig:combined_plots} depicts the variation of MET, $M_{T2}^{Combo}$, and transverse mass of the $b$-jet for the signal (all BPs) and background. With a larger mass gap, the transverse momentum of missing particles, visible lepton and jets increases as they are generated from the cascade decay of $\Psi$. A larger mass gap raises MET, flattening its range and enhancing signal-background separation. A larger mass gap also increases the transverse momentum of the $b$-jet and lepton, along with the missing transverse energy, resulting in an increased transverse mass for the $b$-jet and lepton. This increased transverse mass increases $M_{T2}^{Combo}$, resulting in a more distinct separation between signal and background.
 
%==================
\begin{figure}[tb!]
\centering
  \subfloat {\label{output_bp1}\includegraphics[width=0.43\textwidth]{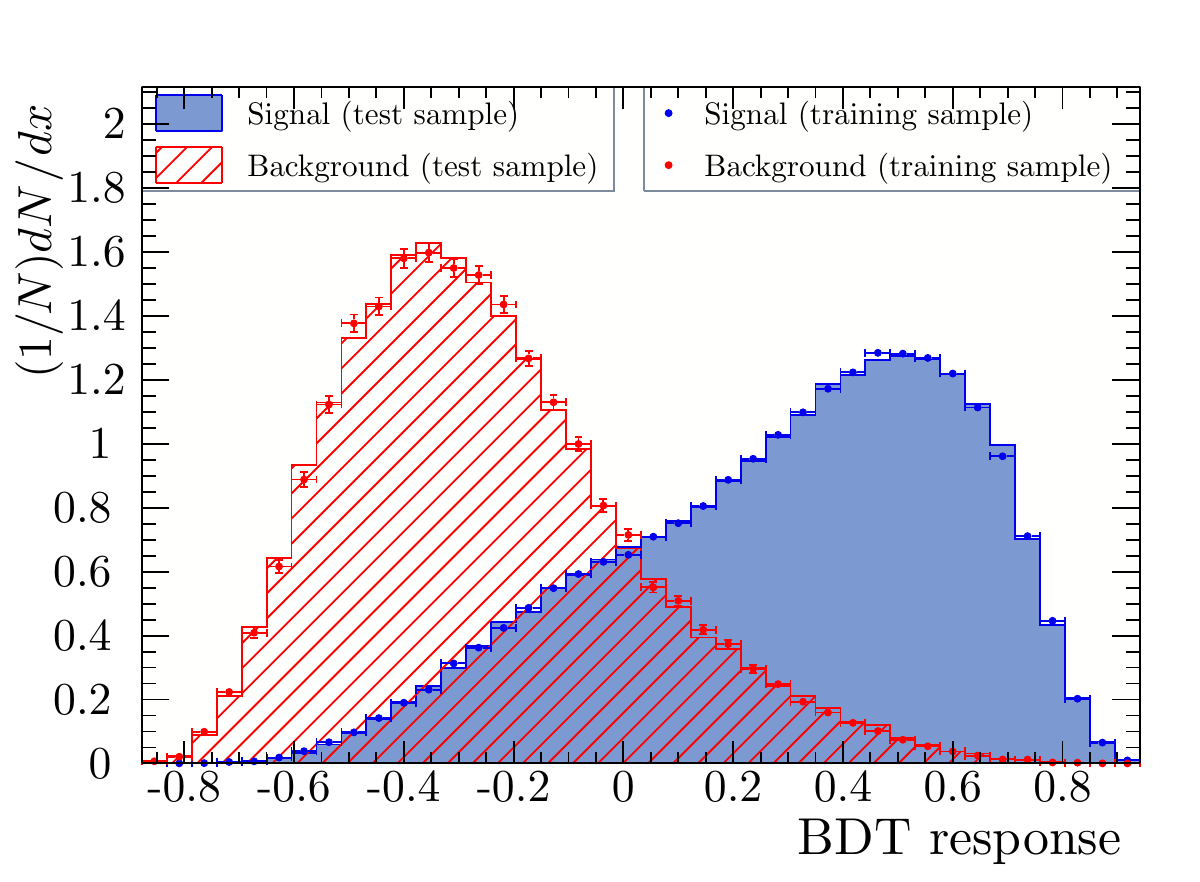}} %\hspace{0.4cm}
  \subfloat {\label{significance_bp1}\includegraphics[width=0.57\textwidth]{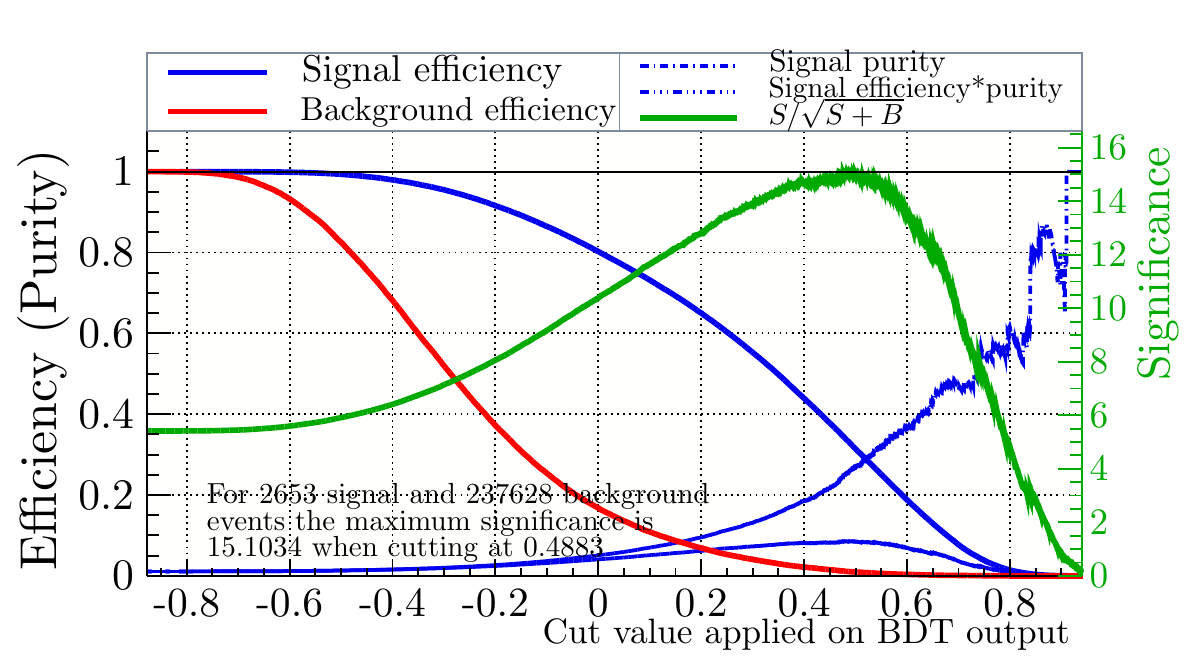}} 
  \caption{On the left, we present the normalized distribution of the BDT response for both signal (BP2) and background classes for training and testing samples. On the right, we display the signal (blue) and background (red) efficiencies alongside the statistical significance of the signal (green), plotted against the cut applied to the BDT output.}\label{MVA:result}
\end{figure}
%==================

%==================
\begin{table}[tb!]
\begin{center}
 \scriptsize
 \begin{tabular}{|c|c|c|c|c|c|c|}
\hline
 & $N_S^{bc}$ & BD$T_{opt}$ & $N_S$ & $N_B$ & $\frac{N_S}{\sqrt{N_S+N_B}}$ ($\frac{N_S}{\sqrt{N_B}})$, 300 $\text{fb}^{-1}$ & $\frac{N_S}{N_B}$  \\ 
\hline\hline
BP1 & 688 & 0.2696 & 209 & 4490 & 3.05 (3.12) & 0.0465 \\
\hline
BP2 & 2653 & 0.4883 & 876 & 2481 & 15.1 (17.6) & 0.353 \\
\hline
BP3 & 590 & 0.4559 & 168 & 367 & 7.26 (8.77) & 0.4578 \\
\hline
BP4 & 342 & 0.4327 & 126 & 243 & 6.56 (8.08) & 0.5185 \\
\hline
BP5 & 35 & 0.6796 & 14 & 75 & 1.48 (1.62) & 0.1867 \\
\hline
$N_{SM}$ & 237628 & \multicolumn{5}{| c |}{} \\
\hline
 \end{tabular} 
\caption{The table displays the effectiveness of the current search in terms of statistical significance for signal over background. $N_S^{bc}$ and $N_{SM}$ represent the total number of events for the combined signal and background before any BDT cuts at a luminosity of 300 $\text{fb}^{-1}$. After optimal selection on the BDT response (BD$T_{opt}$), the surviving signal and background events ($N_S$ and $N_B$) for the 14 TeV LHC are listed. The statistical significance and the signal-to-background ratio are provided in the last two columns.}
\label{tab:BDT}
\end{center}
\end{table}
%==================

Figure \ref{output_bp1} shows the normalized distribution of the BDT response for the signal (BP2) and background classes, including both training and testing samples. The distributions for training and testing match well, and there is a clear separation between the signal and background. In the right panel of figure \ref{MVA:result}, we present the variation in signal (blue) and background (red) efficiencies as well as the statistical significance (green) of the signal compared to the cut applied on the BDT response.

In table \ref{tab:BDT}, the numbers of signal and background events after applying cuts, as discussed below in equation \ref{Eq.BDT inputs}, are represented by $N_S^{bc}$ and $N_{SM}$, respectively, for an integrated luminosity of 300 $\text{fb}^{-1}$. $N_S$, $N_B$ represent the numbers of surviving signal and background events after applying an optimal cut on the BDT output, BD$T_{opt}$. The optimal cut maximizing significance. The table also contains the statistical significance of the signal for different BPs and the signal-to-background ratio. Importantly, for BP1-BP4, we achieve a statistical significance greater than 3.0 for an integrated luminosity of 300 $\text{fb}^{-1}$. For example, in BP4, where the mass of the VLQ is 1.1 TeV, and the mass of the WIMP is 500 GeV, we obtain statistical significance, $\frac{N_S}{\sqrt{N_S+N_B}}=6.56$. 

%======================================================================================
\subsection{Exclusion contour}
%======================================================================================
%==================
\begin{figure}[tb!]
\centering
\includegraphics[width=0.7\textwidth]{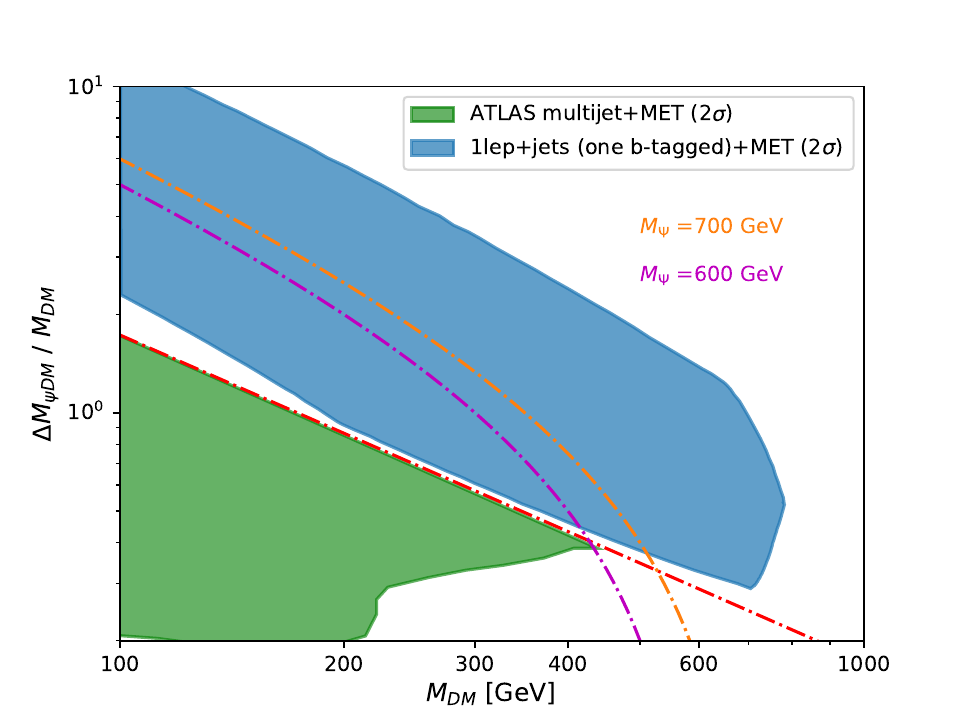}  
\caption{Excluded regions at a $95\%$ CL in the plane of WIMP mass ($M_{\text{DM}}$) and $\frac{\Delta M_{\Psi \text{DM}}}{M_{\text{DM}}}$, where $\Delta M_{\Psi \text{DM}}=M_\Psi - M_{\text{DM}}$, for $\Psi \bar{\Psi}$ production, assuming BR($\Psi\rightarrow t H^{-}$)+BR($\Psi\rightarrow b H^0/A^0) = 1$, and the mass of $\Psi$ is greater than the top-quark mass ($m_t$). Models within the blue region are excluded. The y-axis is plotted on a logarithmic scale. The red dashed line corresponding to $\Delta M_{\Psi \text{DM}}=m_t$ and below this line top is produced in off-shell. The dashed orange and magenta lines represent how a fixed value of $M_{\Psi} = 700~ \text{and}~ 600$ GeV looks in this plane. On the other hand, if $\Psi$ predominantly interacts with the first two generations of the SM quarks, then the exclusion contour at a $95\%$  CL is shown by the filled green region from the reinterpreted analysis \cite{Giacchino:2015hvk} of the ATLAS result \cite{Marjanovic:2014eca}.}
\label{fig:result}
\end{figure}
%==================

We did a detailed collider analysis of this model, and a few representative benchmark points are outlined in the last section. The exclusion contour with a $95\%$ confidence level of our search in the final state of an energetic lepton with multijet where one of the leading two jets is b-tagged and a large missing transverse energy is shown in figure \ref{fig:result}. The exclusion region of our search is shown by a filled blue region in a plane of the dark matter mass, $M_{\text{DM}}$ versus $\frac{\Delta M_{\Psi \text{DM}}}{M_{\text{DM}}}$, where $\Delta M_{\Psi \text{DM}}=M_\Psi - M_{\text{DM}}$ is the mass difference between the vector-like quark ($\Psi$) and DM. The contour is drawn assuming that $\Psi$ primarily interacts with the third generation of the SM quark, with BR($\Psi\rightarrow t H^{-}$)+BR($\Psi\rightarrow b H^0/A^0) = 1$. Also, the mass of $\Psi$ is greater than the top-quark mass, ensuring that the top-quark resulting from the decay of $\Psi$ is on-shell. This contour is plotted for an integrated luminosity of 300 $\text{fb}^{-1}$ at the 14 TeV LHC. Our search strategy is model-independent since we consider $\Psi \bar{\Psi}$ production, where the production cross-section entirely depends on the QCD coupling strength and is independent of any BSM coupling. The y-axis is plotted on a logarithmic scale. We find that a vast parameter space can be probed at the LHC with the current luminosity. From this plot, for instance, if the WIMP mass is 515 GeV (780 GeV), the VLQ mass less than 1.39 TeV (1.19 TeV) can be excluded at $95\%$ CL. 

In the same figure, we present the exclusion contour at a $95\%$  CL (filled green region) if VLQ predominantly interacts with the first two generations of the SM quarks from the reinterpreted analysis \cite{Giacchino:2015hvk} of the ATLAS result \cite{Marjanovic:2014eca}.

%======================================================================================
\section{Summary and conclusion}
\label{conc}
%======================================================================================

The Inert Higgs Doublet Model (IDM) offers a straightforward extension for Weakly Interacting Massive Particle (WIMP) dark matter, typically requiring an ad-hoc discrete symmetry to ensure stability. On the other side, Peccei-Quinn (PQ) symmetry, originally introduced as an elegant solution to the strong CP problem, also provides another compelling dark matter candidate in the form of the axion. In this work, we consider a well-motivated, multicomponent WIMP-axion dark matter model, where the spontaneous breaking of PQ symmetry naturally stabilises the WIMP dark matter through a residual $Z_2$ symmetry. Their mutual contribution ensures the realisation of the observed DM relic density and revives the phenomenologically exciting region of the IDM parameter space where the WIMP mass falls between 100 and 550 GeV. Moreover, additional fields necessitated by PQ symmetry further enrich the dark sector. These include a scalar field proprietor for axion DM and a vector-like quark (VLQ) that acts as a portal between the dark sector and Standard Model particles via Yukawa interactions.

Traditional searches for this degenerate mass region in the IDM rely on displaced vertex searches. The WIMP-axion dark matter model facilitates a more effective, model-independent approach to the Large Hadron Collider (LHC) by exploring the pair production of VLQs. To enhance accuracy, we incorporate Next-to-Leading-Order (NLO) QCD corrections. Following production, the residual $Z_2$ symmetry dictates the decay of each VLQ into a top or bottom quark along with an inert scalar. Employing a leptonic search channel with relevant observables and employing multivariate analysis techniques, we demonstrate the ability to exclude a significant portion of the parameter space using an integrated luminosity of 300 fb$^{-1}$ at the 14 TeV LHC.

In conclusion, this work highlights the potential of the IDM assisted by PQ symmetry as a framework for exploring the WIMP-axion dark matter scenario at the LHC. The proposed search strategy offers a promising avenue for probing previously unexplored regions of the parameter space.

\section*{Acknowledgements}
The computational works are performed using the Param Vikram-1000 High-Performance Computing Cluster and the TDP project resources of the Physical Research Laboratory (PRL).

%%===================================================//
%\appendix
%\section{Appendix}
%\label{appen}
%%===================================================//

%======================================================================================
\bibliographystyle{JHEP}
\bibliography{ref.bib}

\providecommand{\href}[2]{#2}\begingroup\raggedright\begin{thebibliography}{10}

\bibitem{Peccei:1977hh}
R.~D. Peccei and H.~R. Quinn, \emph{{CP Conservation in the Presence of
  Instantons}}, \href{https://doi.org/10.1103/PhysRevLett.38.1440}{\emph{Phys.
  Rev. Lett.} {\bfseries 38} (1977) 1440}.

\bibitem{Peccei:1977ur}
R.~D. Peccei and H.~R. Quinn, \emph{{Constraints Imposed by CP Conservation in
  the Presence of Instantons}},
  \href{https://doi.org/10.1103/PhysRevD.16.1791}{\emph{Phys. Rev. D}
  {\bfseries 16} (1977) 1791}.

\bibitem{Sofue:2000jx}
Y.~Sofue and V.~Rubin, \emph{{Rotation curves of spiral galaxies}},
  \href{https://doi.org/10.1146/annurev.astro.39.1.137}{\emph{Ann. Rev. Astron.
  Astrophys.} {\bfseries 39} (2001) 137}
  [\href{https://arxiv.org/abs/astro-ph/0010594}{{\ttfamily
  astro-ph/0010594}}].

\bibitem{Clowe:2006eq}
D.~Clowe, M.~Bradac, A.~H. Gonzalez, M.~Markevitch, S.~W. Randall, C.~Jones
  et~al., \emph{{A direct empirical proof of the existence of dark matter}},
  \href{https://doi.org/10.1086/508162}{\emph{Astrophys. J. Lett.} {\bfseries
  648} (2006) L109} [\href{https://arxiv.org/abs/astro-ph/0608407}{{\ttfamily
  astro-ph/0608407}}].

\bibitem{Super-Kamiokande:1998kpq}
{\scshape Super-Kamiokande} collaboration, \emph{{Evidence for oscillation of
  atmospheric neutrinos}},
  \href{https://doi.org/10.1103/PhysRevLett.81.1562}{\emph{Phys. Rev. Lett.}
  {\bfseries 81} (1998) 1562}
  [\href{https://arxiv.org/abs/hep-ex/9807003}{{\ttfamily hep-ex/9807003}}].

\bibitem{SNO:2002tuh}
{\scshape SNO} collaboration, \emph{{Direct evidence for neutrino flavor
  transformation from neutral current interactions in the Sudbury Neutrino
  Observatory}},
  \href{https://doi.org/10.1103/PhysRevLett.89.011301}{\emph{Phys. Rev. Lett.}
  {\bfseries 89} (2002) 011301}
  [\href{https://arxiv.org/abs/nucl-ex/0204008}{{\ttfamily nucl-ex/0204008}}].

\bibitem{K2K:2002icj}
{\scshape K2K} collaboration, \emph{{Indications of neutrino oscillation in a
  250 km long baseline experiment}},
  \href{https://doi.org/10.1103/PhysRevLett.90.041801}{\emph{Phys. Rev. Lett.}
  {\bfseries 90} (2003) 041801}
  [\href{https://arxiv.org/abs/hep-ex/0212007}{{\ttfamily hep-ex/0212007}}].

\bibitem{Riotto:1999yt}
A.~Riotto and M.~Trodden, \emph{{Recent progress in baryogenesis}},
  \href{https://doi.org/10.1146/annurev.nucl.49.1.35}{\emph{Ann. Rev. Nucl.
  Part. Sci.} {\bfseries 49} (1999) 35}
  [\href{https://arxiv.org/abs/hep-ph/9901362}{{\ttfamily hep-ph/9901362}}].

\bibitem{Dine:2003ax}
M.~Dine and A.~Kusenko, \emph{{The Origin of the matter - antimatter
  asymmetry}}, \href{https://doi.org/10.1103/RevModPhys.76.1}{\emph{Rev. Mod.
  Phys.} {\bfseries 76} (2003) 1}
  [\href{https://arxiv.org/abs/hep-ph/0303065}{{\ttfamily hep-ph/0303065}}].

\bibitem{Barbieri:2006dq}
R.~Barbieri, L.~J. Hall and V.~S. Rychkov, \emph{{Improved naturalness with a
  heavy Higgs: An Alternative road to LHC physics}},
  \href{https://doi.org/10.1103/PhysRevD.74.015007}{\emph{Phys. Rev. D}
  {\bfseries 74} (2006) 015007}
  [\href{https://arxiv.org/abs/hep-ph/0603188}{{\ttfamily hep-ph/0603188}}].

\bibitem{Cirelli:2005uq}
M.~Cirelli, N.~Fornengo and A.~Strumia, \emph{{Minimal dark matter}},
  \href{https://doi.org/10.1016/j.nuclphysb.2006.07.012}{\emph{Nucl. Phys. B}
  {\bfseries 753} (2006) 178}
  [\href{https://arxiv.org/abs/hep-ph/0512090}{{\ttfamily hep-ph/0512090}}].

\bibitem{Ilnicka:2015jba}
A.~Ilnicka, M.~Krawczyk and T.~Robens, \emph{{Inert Doublet Model in light of
  LHC Run I and astrophysical data}},
  \href{https://doi.org/10.1103/PhysRevD.93.055026}{\emph{Phys. Rev. D}
  {\bfseries 93} (2016) 055026}
  [\href{https://arxiv.org/abs/1508.01671}{{\ttfamily 1508.01671}}].

\bibitem{Belyaev:2016lok}
A.~Belyaev, G.~Cacciapaglia, I.~P. Ivanov, F.~Rojas-Abatte and M.~Thomas,
  \emph{{Anatomy of the Inert Two Higgs Doublet Model in the light of the LHC
  and non-LHC Dark Matter Searches}},
  \href{https://doi.org/10.1103/PhysRevD.97.035011}{\emph{Phys. Rev. D}
  {\bfseries 97} (2018) 035011}
  [\href{https://arxiv.org/abs/1612.00511}{{\ttfamily 1612.00511}}].

\bibitem{Arhrib:2013ela}
A.~Arhrib, Y.-L.~S. Tsai, Q.~Yuan and T.-C. Yuan, \emph{{An Updated Analysis of
  Inert Higgs Doublet Model in light of the Recent Results from LUX, PLANCK,
  AMS-02 and LHC}},
  \href{https://doi.org/10.1088/1475-7516/2014/06/030}{\emph{JCAP} {\bfseries
  06} (2014) 030} [\href{https://arxiv.org/abs/1310.0358}{{\ttfamily
  1310.0358}}].

\bibitem{Dercks:2018wch}
D.~Dercks and T.~Robens, \emph{{Constraining the Inert Doublet Model using
  Vector Boson Fusion}},
  \href{https://doi.org/10.1140/epjc/s10052-019-7436-6}{\emph{Eur. Phys. J. C}
  {\bfseries 79} (2019) 924}
  [\href{https://arxiv.org/abs/1812.07913}{{\ttfamily 1812.07913}}].

\bibitem{Peskin:1991sw}
M.~E. Peskin and T.~Takeuchi, \emph{{Estimation of oblique electroweak
  corrections}}, \href{https://doi.org/10.1103/PhysRevD.46.381}{\emph{Phys.
  Rev. D} {\bfseries 46} (1992) 381}.

\bibitem{Ghosh:2021noq}
A.~Ghosh, P.~Konar and S.~Seth, \emph{{Precise probing of the inert
  Higgs-doublet model at the LHC}},
  \href{https://doi.org/10.1103/PhysRevD.105.115038}{\emph{Phys. Rev. D}
  {\bfseries 105} (2022) 115038}
  [\href{https://arxiv.org/abs/2111.15236}{{\ttfamily 2111.15236}}].

\bibitem{Belyaev:2018ext}
A.~Belyaev, T.~R. Fernandez Perez~Tomei, P.~G. Mercadante, C.~S. Moon,
  S.~Moretti, S.~F. Novaes et~al., \emph{{Advancing LHC probes of dark matter
  from the inert two-Higgs-doublet model with the monojet signal}},
  \href{https://doi.org/10.1103/PhysRevD.99.015011}{\emph{Phys. Rev. D}
  {\bfseries 99} (2019) 015011}
  [\href{https://arxiv.org/abs/1809.00933}{{\ttfamily 1809.00933}}].

\bibitem{Poulose:2016lvz}
P.~Poulose, S.~Sahoo and K.~Sridhar, \emph{{Exploring the Inert Doublet Model
  through the dijet plus missing transverse energy channel at the LHC}},
  \href{https://doi.org/10.1016/j.physletb.2016.12.022}{\emph{Phys. Lett. B}
  {\bfseries 765} (2017) 300}
  [\href{https://arxiv.org/abs/1604.03045}{{\ttfamily 1604.03045}}].

\bibitem{CMS:2014gxa}
{\scshape CMS} collaboration, \emph{{Search for disappearing tracks in
  proton-proton collisions at $ \sqrt{s}=8 $ TeV}},
  \href{https://doi.org/10.1007/JHEP01(2015)096}{\emph{JHEP} {\bfseries 01}
  (2015) 096} [\href{https://arxiv.org/abs/1411.6006}{{\ttfamily 1411.6006}}].

\bibitem{Kim:1979if}
J.~E. Kim, \emph{{Weak Interaction Singlet and Strong CP Invariance}},
  \href{https://doi.org/10.1103/PhysRevLett.43.103}{\emph{Phys. Rev. Lett.}
  {\bfseries 43} (1979) 103}.

\bibitem{Shifman:1979if}
M.~A. Shifman, A.~I. Vainshtein and V.~I. Zakharov, \emph{{Can Confinement
  Ensure Natural CP Invariance of Strong Interactions?}},
  \href{https://doi.org/10.1016/0550-3213(80)90209-6}{\emph{Nucl. Phys. B}
  {\bfseries 166} (1980) 493}.

\bibitem{Dine:1981rt}
M.~Dine, W.~Fischler and M.~Srednicki, \emph{{A Simple Solution to the Strong
  CP Problem with a Harmless Axion}},
  \href{https://doi.org/10.1016/0370-2693(81)90590-6}{\emph{Phys. Lett. B}
  {\bfseries 104} (1981) 199}.

\bibitem{Zhitnitsky:1980tq}
A.~R. Zhitnitsky, \emph{{On Possible Suppression of the Axion Hadron
  Interactions. (In Russian)}}, {\emph{Sov. J. Nucl. Phys.} {\bfseries 31}
  (1980) 260}.

\bibitem{Alves:2016bib}
A.~Alves, D.~A. Camargo, A.~G. Dias, R.~Longas, C.~C. Nishi and F.~S. Queiroz,
  \emph{{Collider and Dark Matter Searches in the Inert Doublet Model from
  Peccei-Quinn Symmetry}},
  \href{https://doi.org/10.1007/JHEP10(2016)015}{\emph{JHEP} {\bfseries 10}
  (2016) 015} [\href{https://arxiv.org/abs/1606.07086}{{\ttfamily
  1606.07086}}].

\bibitem{GrillidiCortona:2015jxo}
G.~Grilli~di Cortona, E.~Hardy, J.~Pardo~Vega and G.~Villadoro, \emph{{The QCD
  axion, precisely}},
  \href{https://doi.org/10.1007/JHEP01(2016)034}{\emph{JHEP} {\bfseries 01}
  (2016) 034} [\href{https://arxiv.org/abs/1511.02867}{{\ttfamily
  1511.02867}}].

\bibitem{Garny:2014waa}
M.~Garny, A.~Ibarra, S.~Rydbeck and S.~Vogl, \emph{{Majorana Dark Matter with a
  Coloured Mediator: Collider vs Direct and Indirect Searches}},
  \href{https://doi.org/10.1007/JHEP06(2014)169}{\emph{JHEP} {\bfseries 06}
  (2014) 169} [\href{https://arxiv.org/abs/1403.4634}{{\ttfamily 1403.4634}}].

\bibitem{Gedalia:2009kh}
O.~Gedalia, Y.~Grossman, Y.~Nir and G.~Perez, \emph{{Lessons from Recent
  Measurements of D0 - anti-D0 Mixing}},
  \href{https://doi.org/10.1103/PhysRevD.80.055024}{\emph{Phys. Rev. D}
  {\bfseries 80} (2009) 055024}
  [\href{https://arxiv.org/abs/0906.1879}{{\ttfamily 0906.1879}}].

\bibitem{Chatterjee:2018mac}
S.~Chatterjee, A.~Das, T.~Samui and M.~Sen, \emph{{Mixed WIMP-axion dark
  matter}}, \href{https://doi.org/10.1103/PhysRevD.100.115050}{\emph{Phys. Rev.
  D} {\bfseries 100} (2019) 115050}
  [\href{https://arxiv.org/abs/1810.09471}{{\ttfamily 1810.09471}}].

\bibitem{Deshpande:1977rw}
N.~G. Deshpande and E.~Ma, \emph{{Pattern of Symmetry Breaking with Two Higgs
  Doublets}}, \href{https://doi.org/10.1103/PhysRevD.18.2574}{\emph{Phys. Rev.
  D} {\bfseries 18} (1978) 2574}.

\bibitem{Ivanov:2006yq}
I.~P. Ivanov, \emph{{Minkowski space structure of the Higgs potential in
  2HDM}}, \href{https://doi.org/10.1103/PhysRevD.75.035001}{\emph{Phys. Rev. D}
  {\bfseries 75} (2007) 035001}
  [\href{https://arxiv.org/abs/hep-ph/0609018}{{\ttfamily hep-ph/0609018}}].

\bibitem{Ginzburg:2010wa}
I.~F. Ginzburg, K.~A. Kanishev, M.~Krawczyk and D.~Sokolowska, \emph{{Evolution
  of Universe to the present inert phase}},
  \href{https://doi.org/10.1103/PhysRevD.82.123533}{\emph{Phys. Rev. D}
  {\bfseries 82} (2010) 123533}
  [\href{https://arxiv.org/abs/1009.4593}{{\ttfamily 1009.4593}}].

\bibitem{Swiezewska:2012ej}
B.~\'Swie\.zewska, \emph{{Yukawa independent constraints for two-Higgs-doublet
  models with a 125 GeV Higgs boson}},
  \href{https://doi.org/10.1103/PhysRevD.88.055027}{\emph{Phys. Rev. D}
  {\bfseries 88} (2013) 055027}
  [\href{https://arxiv.org/abs/1209.5725}{{\ttfamily 1209.5725}}].

\bibitem{Arhrib:2012ia}
A.~Arhrib, R.~Benbrik and N.~Gaur, \emph{{$H\to \gamma \gamma$ in Inert Higgs
  Doublet Model}},
  \href{https://doi.org/10.1103/PhysRevD.85.095021}{\emph{Phys. Rev. D}
  {\bfseries 85} (2012) 095021}
  [\href{https://arxiv.org/abs/1201.2644}{{\ttfamily 1201.2644}}].

\bibitem{Belanger:2013xza}
G.~Belanger, B.~Dumont, U.~Ellwanger, J.~F. Gunion and S.~Kraml, \emph{{Global
  fit to Higgs signal strengths and couplings and implications for extended
  Higgs sectors}},
  \href{https://doi.org/10.1103/PhysRevD.88.075008}{\emph{Phys. Rev. D}
  {\bfseries 88} (2013) 075008}
  [\href{https://arxiv.org/abs/1306.2941}{{\ttfamily 1306.2941}}].

\bibitem{Lundstrom:2008ai}
E.~Lundstrom, M.~Gustafsson and J.~Edsjo, \emph{{The Inert Doublet Model and
  LEP II Limits}},
  \href{https://doi.org/10.1103/PhysRevD.79.035013}{\emph{Phys. Rev. D}
  {\bfseries 79} (2009) 035013}
  [\href{https://arxiv.org/abs/0810.3924}{{\ttfamily 0810.3924}}].

\bibitem{Belanger:2015kga}
G.~Belanger, B.~Dumont, A.~Goudelis, B.~Herrmann, S.~Kraml and D.~Sengupta,
  \emph{{Dilepton constraints in the Inert Doublet Model from Run 1 of the
  LHC}}, \href{https://doi.org/10.1103/PhysRevD.91.115011}{\emph{Phys. Rev. D}
  {\bfseries 91} (2015) 115011}
  [\href{https://arxiv.org/abs/1503.07367}{{\ttfamily 1503.07367}}].

\bibitem{Pierce:2007ut}
A.~Pierce and J.~Thaler, \emph{{Natural Dark Matter from an Unnatural Higgs
  Boson and New Colored Particles at the TeV Scale}},
  \href{https://doi.org/10.1088/1126-6708/2007/08/026}{\emph{JHEP} {\bfseries
  08} (2007) 026} [\href{https://arxiv.org/abs/hep-ph/0703056}{{\ttfamily
  hep-ph/0703056}}].

\bibitem{Marjanovic:2014eca}
{\scshape ATLAS} collaboration, \emph{{Search for squarks and gluinos with the
  ATLAS detector in final states with jets and missing transverse momentum
  using 20.3 $fb^{-1}$ of $\sqrt{s}$ = 8 TeV proton-proton collision data}},
  in \emph{{2nd Large Hadron Collider Physics Conference}}, 8, 2014,
  \href{https://arxiv.org/abs/1408.5857}{{\ttfamily 1408.5857}}.

\bibitem{Giacchino:2015hvk}
F.~Giacchino, A.~Ibarra, L.~Lopez~Honorez, M.~H.~G. Tytgat and S.~Wild,
  \emph{{Signatures from Scalar Dark Matter with a Vector-like Quark
  Mediator}}, \href{https://doi.org/10.1088/1475-7516/2016/02/002}{\emph{JCAP}
  {\bfseries 02} (2016) 002}
  [\href{https://arxiv.org/abs/1511.04452}{{\ttfamily 1511.04452}}].

\bibitem{LUX:2015abn}
{\scshape LUX} collaboration, \emph{{Improved Limits on Scattering of Weakly
  Interacting Massive Particles from Reanalysis of 2013 LUX Data}},
  \href{https://doi.org/10.1103/PhysRevLett.116.161301}{\emph{Phys. Rev. Lett.}
  {\bfseries 116} (2016) 161301}
  [\href{https://arxiv.org/abs/1512.03506}{{\ttfamily 1512.03506}}].

\bibitem{XENON100:2012itz}
{\scshape XENON100} collaboration, \emph{{Dark Matter Results from 225 Live
  Days of XENON100 Data}},
  \href{https://doi.org/10.1103/PhysRevLett.109.181301}{\emph{Phys. Rev. Lett.}
  {\bfseries 109} (2012) 181301}
  [\href{https://arxiv.org/abs/1207.5988}{{\ttfamily 1207.5988}}].

\bibitem{SuperCDMS:2014cds}
{\scshape SuperCDMS} collaboration, \emph{{Search for Low-Mass Weakly
  Interacting Massive Particles with SuperCDMS}},
  \href{https://doi.org/10.1103/PhysRevLett.112.241302}{\emph{Phys. Rev. Lett.}
  {\bfseries 112} (2014) 241302}
  [\href{https://arxiv.org/abs/1402.7137}{{\ttfamily 1402.7137}}].

\bibitem{CDMS:2008uih}
{\scshape CDMS} collaboration, \emph{{Search for Weakly Interacting Massive
  Particles with the First Five-Tower Data from the Cryogenic Dark Matter
  Search at the Soudan Underground Laboratory}},
  \href{https://doi.org/10.1103/PhysRevLett.102.011301}{\emph{Phys. Rev. Lett.}
  {\bfseries 102} (2009) 011301}
  [\href{https://arxiv.org/abs/0802.3530}{{\ttfamily 0802.3530}}].

\bibitem{Alner:2007ja}
G.~J. Alner et~al., \emph{{First limits on WIMP nuclear recoil signals in
  ZEPLIN-II: A two phase xenon detector for dark matter detection}},
  \href{https://doi.org/10.1016/j.astropartphys.2007.06.002}{\emph{Astropart.
  Phys.} {\bfseries 28} (2007) 287}
  [\href{https://arxiv.org/abs/astro-ph/0701858}{{\ttfamily
  astro-ph/0701858}}].

\bibitem{T.Hambye_2009}
T.~Hambye, F.-S. Ling, L.~L. Honorez and J.~Rocher, \emph{Scalar multiplet dark
  matter}, \href{https://doi.org/10.1088/1126-6708/2009/07/090}{\emph{Journal
  of High Energy Physics} {\bfseries 2009} (2009) 090}.

\bibitem{Blinov:2015qva}
N.~Blinov, J.~Kozaczuk, D.~E. Morrissey and A.~de~la Puente, \emph{{Compressing
  the Inert Doublet Model}},
  \href{https://doi.org/10.1103/PhysRevD.93.035020}{\emph{Phys. Rev. D}
  {\bfseries 93} (2016) 035020}
  [\href{https://arxiv.org/abs/1510.08069}{{\ttfamily 1510.08069}}].

\bibitem{Aprile:2012zx}
{\scshape XENON1T} collaboration, \emph{{The XENON1T Dark Matter Search
  Experiment}},
  \href{https://doi.org/10.1007/978-94-007-7241-0_14}{\emph{Springer Proc.
  Phys.} {\bfseries 148} (2013) 93}
  [\href{https://arxiv.org/abs/1206.6288}{{\ttfamily 1206.6288}}].

\bibitem{XENON:2015gkh}
{\scshape XENON} collaboration, \emph{{Physics reach of the XENON1T dark matter
  experiment}},
  \href{https://doi.org/10.1088/1475-7516/2016/04/027}{\emph{JCAP} {\bfseries
  04} (2016) 027} [\href{https://arxiv.org/abs/1512.07501}{{\ttfamily
  1512.07501}}].

\bibitem{Kohri:2009yn}
K.~Kohri, A.~Mazumdar, N.~Sahu and P.~Stephens, \emph{{Probing Unified Origin
  of Dark Matter and Baryon Asymmetry at PAMELA/Fermi}},
  \href{https://doi.org/10.1103/PhysRevD.80.061302}{\emph{Phys. Rev. D}
  {\bfseries 80} (2009) 061302}
  [\href{https://arxiv.org/abs/0907.0622}{{\ttfamily 0907.0622}}].

\bibitem{MAGIC:2016xys}
{\scshape MAGIC, Fermi-LAT} collaboration, \emph{{Limits to Dark Matter
  Annihilation Cross-Section from a Combined Analysis of MAGIC and Fermi-LAT
  Observations of Dwarf Satellite Galaxies}},
  \href{https://doi.org/10.1088/1475-7516/2016/02/039}{\emph{JCAP} {\bfseries
  02} (2016) 039} [\href{https://arxiv.org/abs/1601.06590}{{\ttfamily
  1601.06590}}].

\bibitem{Eiteneuer:2017hoh}
B.~Eiteneuer, A.~Goudelis and J.~Heisig, \emph{{The inert doublet model in the
  light of Fermi-LAT gamma-ray data: a global fit analysis}},
  \href{https://doi.org/10.1140/epjc/s10052-017-5166-1}{\emph{Eur. Phys. J. C}
  {\bfseries 77} (2017) 624}
  [\href{https://arxiv.org/abs/1705.01458}{{\ttfamily 1705.01458}}].

\bibitem{HESS:2011zpk}
{\scshape H.E.S.S.} collaboration, \emph{{Search for a Dark Matter annihilation
  signal from the Galactic Center halo with H.E.S.S}},
  \href{https://doi.org/10.1103/PhysRevLett.106.161301}{\emph{Phys. Rev. Lett.}
  {\bfseries 106} (2011) 161301}
  [\href{https://arxiv.org/abs/1103.3266}{{\ttfamily 1103.3266}}].

\bibitem{Aghanim:2018eyx}
{\scshape Planck} collaboration, \emph{{Planck 2018 results. VI. Cosmological
  parameters}},  \href{https://arxiv.org/abs/1807.06209}{{\ttfamily
  1807.06209}}.

\bibitem{Belanger:2018ccd}
G.~B\'elanger, F.~Boudjema, A.~Goudelis, A.~Pukhov and B.~Zaldivar,
  \emph{{micrOMEGAs5.0 : Freeze-in}},
  \href{https://doi.org/10.1016/j.cpc.2018.04.027}{\emph{Comput. Phys. Commun.}
  {\bfseries 231} (2018) 173}
  [\href{https://arxiv.org/abs/1801.03509}{{\ttfamily 1801.03509}}].

\bibitem{Datta:2016nfz}
A.~Datta, N.~Ganguly, N.~Khan and S.~Rakshit, \emph{{Exploring collider
  signatures of the inert Higgs doublet model}},
  \href{https://doi.org/10.1103/PhysRevD.95.015017}{\emph{Phys. Rev. D}
  {\bfseries 95} (2017) 015017}
  [\href{https://arxiv.org/abs/1610.00648}{{\ttfamily 1610.00648}}].

\bibitem{Berezhiani:1989fp}
Z.~G. Berezhiani and M.~Y. Khlopov, \emph{{Cosmology of Spontaneously Broken
  Gauge Family Symmetry}}, \href{https://doi.org/10.1007/BF01570798}{\emph{Z.
  Phys. C} {\bfseries 49} (1991) 73}.

\bibitem{Ghosh:2023xhs}
A.~Ghosh and P.~Konar, \emph{{Precision prediction at the LHC of a democratic
  up-family philic KSVZ axion model}},
  \href{https://arxiv.org/abs/2305.08662}{{\ttfamily 2305.08662}}.

\bibitem{Sikivie:2006ni}
P.~Sikivie, \emph{{Axion Cosmology}},
  \href{https://doi.org/10.1007/978-3-540-73518-2_2}{\emph{Lect. Notes Phys.}
  {\bfseries 741} (2008) 19}
  [\href{https://arxiv.org/abs/astro-ph/0610440}{{\ttfamily
  astro-ph/0610440}}].

\bibitem{Bae:2008ue}
K.~J. Bae, J.-H. Huh and J.~E. Kim, \emph{{Update of axion CDM energy}},
  \href{https://doi.org/10.1088/1475-7516/2008/09/005}{\emph{JCAP} {\bfseries
  09} (2008) 005} [\href{https://arxiv.org/abs/0806.0497}{{\ttfamily
  0806.0497}}].

\bibitem{Ghosh:2023ocz}
A.~Ghosh, P.~Konar, D.~Saha and S.~Seth, \emph{{Precise probing and
  discrimination of third-generation scalar leptoquarks}},
  \href{https://doi.org/10.1103/PhysRevD.108.035030}{\emph{Phys. Rev. D}
  {\bfseries 108} (2023) 035030}
  [\href{https://arxiv.org/abs/2304.02890}{{\ttfamily 2304.02890}}].

\bibitem{Ghosh:2022rta}
A.~Ghosh, P.~Konar and R.~Roshan, \emph{{Top-philic dark matter in a hybrid
  KSVZ axion framework}},
  \href{https://doi.org/10.1007/JHEP12(2022)167}{\emph{JHEP} {\bfseries 12}
  (2022) 167} [\href{https://arxiv.org/abs/2207.00487}{{\ttfamily
  2207.00487}}].

\bibitem{Mangano:2006rw}
M.~L. Mangano, M.~Moretti, F.~Piccinini and M.~Treccani, \emph{{Matching matrix
  elements and shower evolution for top-quark production in hadronic
  collisions}},
  \href{https://doi.org/10.1088/1126-6708/2007/01/013}{\emph{JHEP} {\bfseries
  01} (2007) 013} [\href{https://arxiv.org/abs/hep-ph/0611129}{{\ttfamily
  hep-ph/0611129}}].

\bibitem{Hoeche:2005vzu}
S.~Hoeche, F.~Krauss, N.~Lavesson, L.~Lonnblad, M.~Mangano, A.~Schalicke
  et~al., \emph{{Matching parton showers and matrix elements}},  in \emph{{HERA
  and the LHC: A Workshop on the Implications of HERA for LHC Physics: CERN -
  DESY Workshop 2004/2005 (Midterm Meeting, CERN, 11-13 October 2004; Final
  Meeting, DESY, 17-21 January 2005)}}, pp.~288--289, 2005,
  \href{https://arxiv.org/abs/hep-ph/0602031}{{\ttfamily hep-ph/0602031}},
  \href{https://doi.org/10.5170/CERN-2005-014.288}{DOI}.

\bibitem{Alloul:2013bka}
A.~Alloul, N.~D. Christensen, C.~Degrande, C.~Duhr and B.~Fuks,
  \emph{{FeynRules 2.0 - A complete toolbox for tree-level phenomenology}},
  \href{https://doi.org/10.1016/j.cpc.2014.04.012}{\emph{Comput. Phys. Commun.}
  {\bfseries 185} (2014) 2250}
  [\href{https://arxiv.org/abs/1310.1921}{{\ttfamily 1310.1921}}].

\bibitem{Alwall:2014hca}
J.~Alwall, R.~Frederix, S.~Frixione, V.~Hirschi, F.~Maltoni, O.~Mattelaer
  et~al., \emph{{The automated computation of tree-level and next-to-leading
  order differential cross sections, and their matching to parton shower
  simulations}}, \href{https://doi.org/10.1007/JHEP07(2014)079}{\emph{JHEP}
  {\bfseries 07} (2014) 079} [\href{https://arxiv.org/abs/1405.0301}{{\ttfamily
  1405.0301}}].

\bibitem{Sjostrand:2001yu}
T.~Sjostrand, L.~Lonnblad and S.~Mrenna, \emph{{PYTHIA 6.2: Physics and
  manual}},  \href{https://arxiv.org/abs/hep-ph/0108264}{{\ttfamily
  hep-ph/0108264}}.

\bibitem{Sjostrand:2014zea}
T.~Sj\"ostrand, S.~Ask, J.~R. Christiansen, R.~Corke, N.~Desai, P.~Ilten
  et~al., \emph{{An introduction to PYTHIA 8.2}},
  \href{https://doi.org/10.1016/j.cpc.2015.01.024}{\emph{Comput. Phys. Commun.}
  {\bfseries 191} (2015) 159}
  [\href{https://arxiv.org/abs/1410.3012}{{\ttfamily 1410.3012}}].

\bibitem{deFavereau:2013fsa}
{\scshape DELPHES 3} collaboration, \emph{{DELPHES 3, A modular framework for
  fast simulation of a generic collider experiment}},
  \href{https://doi.org/10.1007/JHEP02(2014)057}{\emph{JHEP} {\bfseries 02}
  (2014) 057} [\href{https://arxiv.org/abs/1307.6346}{{\ttfamily 1307.6346}}].

\bibitem{NNPDF:2014otw}
{\scshape NNPDF} collaboration, \emph{{Parton distributions for the LHC Run
  II}}, \href{https://doi.org/10.1007/JHEP04(2015)040}{\emph{JHEP} {\bfseries
  04} (2015) 040} [\href{https://arxiv.org/abs/1410.8849}{{\ttfamily
  1410.8849}}].

\bibitem{Cacciari:2008gp}
M.~Cacciari, G.~P. Salam and G.~Soyez, \emph{{The anti-$k_t$ jet clustering
  algorithm}}, \href{https://doi.org/10.1088/1126-6708/2008/04/063}{\emph{JHEP}
  {\bfseries 04} (2008) 063} [\href{https://arxiv.org/abs/0802.1189}{{\ttfamily
  0802.1189}}].

\bibitem{Roe:2004na}
B.~P. Roe, H.-J. Yang, J.~Zhu, Y.~Liu, I.~Stancu and G.~McGregor,
  \emph{{Boosted decision trees, an alternative to artificial neural
  networks}}, \href{https://doi.org/10.1016/j.nima.2004.12.018}{\emph{Nucl.
  Instrum. Meth. A} {\bfseries 543} (2005) 577}
  [\href{https://arxiv.org/abs/physics/0408124}{{\ttfamily physics/0408124}}].

\bibitem{FREUND1995256}
Y.~Freund, \emph{Boosting a weak learning algorithm by majority},
  \href{https://doi.org/https://doi.org/10.1006/inco.1995.1136}{\emph{Information
  and Computation} {\bfseries 121} (1995) 256}.

\bibitem{Freund:1997xna}
Y.~Freund and R.~E. Schapire, \emph{{A Decision-Theoretic Generalization of
  On-Line Learning and an Application to Boosting}},
  \href{https://doi.org/10.1006/jcss.1997.1504}{\emph{J. Comput. Syst. Sci.}
  {\bfseries 55} (1997) 119}.

\bibitem{Hocker:2007ht}
A.~Hocker et~al., \emph{{TMVA - Toolkit for Multivariate Data Analysis}},
  \href{https://arxiv.org/abs/physics/0703039}{{\ttfamily physics/0703039}}.

\bibitem{Muselli:2015kba}
C.~Muselli, M.~Bonvini, S.~Forte, S.~Marzani and G.~Ridolfi, \emph{{Top Quark
  Pair Production beyond NNLO}},
  \href{https://doi.org/10.1007/JHEP08(2015)076}{\emph{JHEP} {\bfseries 08}
  (2015) 076} [\href{https://arxiv.org/abs/1505.02006}{{\ttfamily
  1505.02006}}].

\bibitem{Kant:2014oha}
P.~Kant, O.~M. Kind, T.~Kintscher, T.~Lohse, T.~Martini, S.~M\"olbitz et~al.,
  \emph{{HatHor for single top-quark production: Updated predictions and
  uncertainty estimates for single top-quark production in hadronic
  collisions}}, \href{https://doi.org/10.1016/j.cpc.2015.02.001}{\emph{Comput.
  Phys. Commun.} {\bfseries 191} (2015) 74}
  [\href{https://arxiv.org/abs/1406.4403}{{\ttfamily 1406.4403}}].

\bibitem{Kidonakis:2015nna}
N.~Kidonakis, \emph{{Theoretical results for electroweak-boson and single-top
  production}}, \href{https://doi.org/10.22323/1.247.0170}{\emph{PoS}
  {\bfseries DIS2015} (2015) 170}
  [\href{https://arxiv.org/abs/1506.04072}{{\ttfamily 1506.04072}}].

\bibitem{Balossini:2009sa}
G.~Balossini, G.~Montagna, C.~M. Carloni~Calame, M.~Moretti, O.~Nicrosini,
  F.~Piccinini et~al., \emph{{Combination of electroweak and QCD corrections to
  single W production at the Fermilab Tevatron and the CERN LHC}},
  \href{https://doi.org/10.1007/JHEP01(2010)013}{\emph{JHEP} {\bfseries 01}
  (2010) 013} [\href{https://arxiv.org/abs/0907.0276}{{\ttfamily 0907.0276}}].

\bibitem{Campbell:2011bn}
J.~M. Campbell, R.~K. Ellis and C.~Williams, \emph{{Vector boson pair
  production at the LHC}},
  \href{https://doi.org/10.1007/JHEP07(2011)018}{\emph{JHEP} {\bfseries 07}
  (2011) 018} [\href{https://arxiv.org/abs/1105.0020}{{\ttfamily 1105.0020}}].

\bibitem{CMS:2017yfk}
{\scshape CMS} collaboration, \emph{{Particle-flow reconstruction and global
  event description with the CMS detector}},
  \href{https://doi.org/10.1088/1748-0221/12/10/P10003}{\emph{JINST} {\bfseries
  12} (2017) P10003} [\href{https://arxiv.org/abs/1706.04965}{{\ttfamily
  1706.04965}}].

\bibitem{Cheng:2008hk}
H.-C. Cheng and Z.~Han, \emph{{Minimal Kinematic Constraints and m(T2)}},
  \href{https://doi.org/10.1088/1126-6708/2008/12/063}{\emph{JHEP} {\bfseries
  12} (2008) 063} [\href{https://arxiv.org/abs/0810.5178}{{\ttfamily
  0810.5178}}].

\bibitem{Lester:1999tx}
C.~G. Lester and D.~J. Summers, \emph{{Measuring masses of semiinvisibly
  decaying particles pair produced at hadron colliders}},
  \href{https://doi.org/10.1016/S0370-2693(99)00945-4}{\emph{Phys. Lett. B}
  {\bfseries 463} (1999) 99}
  [\href{https://arxiv.org/abs/hep-ph/9906349}{{\ttfamily hep-ph/9906349}}].

\bibitem{Barr:2011xt}
A.~J. Barr, T.~J. Khoo, P.~Konar, K.~Kong, C.~G. Lester, K.~T. Matchev et~al.,
  \emph{{Guide to transverse projections and mass-constraining variables}},
  \href{https://doi.org/10.1103/PhysRevD.84.095031}{\emph{Phys. Rev. D}
  {\bfseries 84} (2011) 095031}
  [\href{https://arxiv.org/abs/1105.2977}{{\ttfamily 1105.2977}}].

\bibitem{Konar:2009wn}
P.~Konar, K.~Kong, K.~T. Matchev and M.~Park, \emph{{Superpartner Mass
  Measurement Technique using 1D Orthogonal Decompositions of the Cambridge
  Transverse Mass Variable $M_{T2}$}},
  \href{https://doi.org/10.1103/PhysRevLett.105.051802}{\emph{Phys. Rev. Lett.}
  {\bfseries 105} (2010) 051802}
  [\href{https://arxiv.org/abs/0910.3679}{{\ttfamily 0910.3679}}].

\bibitem{Konar:2008ei}
P.~Konar, K.~Kong and K.~T. Matchev, \emph{{$\sqrt{\hat{s}}_{min}$ : A Global
  inclusive variable for determining the mass scale of new physics in events
  with missing energy at hadron colliders}},
  \href{https://doi.org/10.1088/1126-6708/2009/03/085}{\emph{JHEP} {\bfseries
  03} (2009) 085} [\href{https://arxiv.org/abs/0812.1042}{{\ttfamily
  0812.1042}}].

\end{thebibliography}\endgroup
%======================================================================================
\end{document}